\documentclass[useAMS,fleqn,usenatbib]{mnras}

\usepackage[T1]{fontenc}
\usepackage{ae,aecompl}

\usepackage{amsmath}
\usepackage{amsopn}
\usepackage[british]{babel}
\usepackage[varg]{txfonts}
\usepackage{biblio} 
\usepackage{natbib}
\usepackage{color}
\usepackage{bm}

\usepackage{graphicx}
\usepackage{epstopdf} 

\usepackage{microtype}

\definecolor{orange}{rgb}{1.0,0.5,0.}

\DeclareMathOperator{\sech}{sech}

\def\MDM{\ifmmode{\>M_{\textnormal{\sc dm}}}\else{$$M_{\textnormal{\sc dm}}}\fi}

\def\XH{\ifmmode{\>X_{\textnormal{\sc h}}} \else{$X_{\textnormal{\sc h}}$}\fi}
\def\nH{\ifmmode{\>n_{\textnormal{\sc h}}} \else{$n_{\textnormal{\sc h}}$}\fi}

\def\maspyr{\ifmmode{\>\textnormal{mas~yr}^{-1}}\else{mas~yr$^{-1}$}\fi}

\def\mG{\ifmmode{\>\mu\mathrm{G}}\else{$\mu$G}\fi}
\def\erg{\ifmmode{\> {\rm erg}}\else{erg}\fi}
\def\keV{\ifmmode{\> {\rm keV}}\else{keV}\fi}

\def\deg{\ifmmode{\>^{\circ}}\else{$^{\circ}$}\fi}
\def\onedeg{\ifmmode{\>1^{\circ}}\else{$1^{\circ}$}\fi}

\def\xvir{\ifmmode{\>\!x_{vir}}\else{$x_{vir}$}\fi}
\def\Mvir{\ifmmode{\>\!M_{vir} }\else{$M_{vir} $}\fi}
\def\rvir{\ifmmode{\>\!r_{vir}}\else{$r_{vir}$}\fi}
\def\vvir{\ifmmode{\>\!v_{vir}}\else{$v_{vir}$}\fi}
\def\Vvir{\ifmmode{\>\!V_{vir} }\else{$V_{vir} $}\fi}

\def\tratio{\ifmmode{\>\tau}\else{$\tau$}\fi}

\def\rms{\ifmmode{\>r_{\textnormal{\sc ms}}}\else{$r_{\textnormal{\sc ms}}$}\fi}

\def\Mpc{\ifmmode{\>\!{\rm Mpc}} \else{Mpc}\fi}
\def\kpc{\ifmmode{\>\!{\rm kpc}} \else{kpc}\fi}
\def\pc{\ifmmode{\>\!{\rm pc}} \else{pc}\fi}

\def\Gyr{\ifmmode{\>\!{\rm Gyr}} \else{Gyr}\fi}
\def\Myr{\ifmmode{\>\!{\rm Myr}} \else{Myr}\fi}
\def\yr{\ifmmode{\>\!{\rm yr}} \else{yr}\fi}
\def\pyr{\ifmmode{\>\!{\rm yr}^{-1}}\else{yr $^{-1}$} \fi}
\def\s{\ifmmode{\>\!{\rm s}}\else{s}\fi}
\def\ps{\ifmmode{\>\!{\rm s}^{-1}}\else{s$^{-1}$}\fi}
\def\Hz{\ifmmode{\>\!{\rm Hz}}\else{Hz}\fi}

\def\kms{\ifmmode{\>\!{\rm km\,s}^{-1}}\else{km~s$^{-1}$}\fi}

\def\K{\ifmmode{\>\!{\rm K}}\else{K}\fi}

\def\sr{\ifmmode{\>\!{\rm sr}}\else{sr}\fi}
\def\psr{\ifmmode{\>\!{\rm sr}^{-1}}\else{sr$^{-1}$}\fi}
\def\arcs{\ifmmode{\>\!{\rm arcsec}}\else{arcsec}\fi}
\def\parcs{\ifmmode{\>\!{\rm arcsec}^{-1}}\else{arcsec${-1}$}\fi}
\def\parcss{\ifmmode{\>\!{\rm arcsec}^{-2}}\else{arcsec${-2}$}\fi}

\def\cm{\ifmmode{\>\!{\rm cm}}\else{cm}\fi}
\def\cc{\ifmmode{\>\!{\rm cm}^{3}}\else{cm$^{3}$}\fi}
\def\sqc{\ifmmode{\>\!{\rm cm}^{2}}\else{cm$^{2}$}\fi}
\def\pcc{\ifmmode{\>\!{\rm cm}^{-3}}\else{cm$^{-3}$}\fi}
\def\psc{\ifmmode{\>\!{\rm cm}^{-2}}\else{cm$^{-2}$}\fi}

\def\g{\ifmmode{\>\!{\rm g}}\else{g}\fi}
\def\Msun{\ifmmode{\>\!{\rm M}_{\odot}}\else{M$_{\odot}$}\fi}
\def\hMsun{\ifmmode{\> h^{-1}{\rm M}_{\odot}}\else{$h^{-1}$M$_{\odot}$}\fi}

\def\Zsun{\ifmmode{\>\!{\rm Z}_{\odot}}\else{Z$_{\odot}$}\fi}

\def\Lsun{\ifmmode{\>\!{\rm L}_{\odot}}\else{L$_{\odot}$}\fi}

\def\rayl{\ifmmode{\>\!{\rm R}}\else{R}\fi}
\def\mR{\ifmmode{\>\!{\rm mR}}\else{mR}\fi}

\renewcommand{\ion}[2]{\hbox{#1\,{\sc #2}}}

\def\lya{\ifmmode{\>\!{\rm Ly}\alpha}\else{Ly$\alpha$}\fi}

\def\Ha{\ifmmode{\>\!{\rm H}\alpha}\else{H$\alpha$}\fi}
\def\Hb{\ifmmode{\>\!{\rm H}\beta}\else{H$\beta$}\fi}

\def\HI{\ifmmode{\> \textnormal{\ion{H}{i}}} \else{\ion{H}{i}}\fi}
\def\HII{\ifmmode{\> \textnormal{\ion{H}{ii}}} \else{\ion{H}{ii}}\fi}
\def\CIV{\ifmmode{\> \textnormal{\ion{C}{iv}}} \else{\ion{C}{iv}}\fi}
\def\SiIV{\ifmmode{\> \textnormal{\ion{S}{iv}}} \else{\ion{Si}{iv}}\fi}

\def\NH{\ifmmode{\> {\rm N}_{\rm H}} \else{N$_{\rm H}$}\fi}
\def\Ng{\ifmmode{\> {\rm N}_{\rm gas}} \else{N$_{\rm gas}$}\fi}
\def\NHI{\ifmmode{\> {\rm N}_{\HI}} \else{N$_{\HI}$}\fi}
\def\MHI{\ifmmode{\> {\rm M}_{ \HI}} \else{M$_{\HI}$}\fi}

\def\mua{\ifmmode{\>\mu_{ \textnormal{\Ha}}}\else{$\mu_{ \textnormal{\Ha}}$}\fi}
\def\alphabha{\ifmmode{\>\alpha_{B}^{(\textnormal{\Ha})}}\else{$\alpha_{B}^{(\textnormal{\Ha})}$}\fi}

\newcommand{\myemail}{tepper@physics.usyd.edu.au}
\newcommand{\ramses}{{\sc Ramses}}
\newcommand{\agama}{{\small AGAMA}}
\newcommand{\pynbody}{{\sc Pynbody}}
\newcommand{\galactics}{{\small GALACTICS}}
\newcommand{\gaia}{{\em Gaia}}
\newcommand{\dice}{{\small DICE}}
\newcommand{\nexus}{{\sc Nexus}}

\title[ {\nexus} ]{ {\sc \textbf{Nexus}}: A framework for controlled simulations of idealised galaxies}

\author[Tepper-Garc\'ia et al.]{%
Thor Tepper-Garc\'ia,$^{1,2}$\thanks{\myemail}
Joss Bland-Hawthorn,$^{1,2}$
Eugene Vasiliev,$^{3,4}$
Oscar Agertz,$^5$
Romain Teyssier,$^6$
\newauthor
and Christoph Federrath$^{7,2}$
\\
$^1$Sydney Institute for Astronomy, School of Physics, University of Sydney, NSW 2006, Australia\\
$^2$Centre of Excellence for All Sky Astrophysics in Three Dimensions (ASTRO-3D), Australia\\
$^3$University of Surrey, Guildford, GU2 7XH, UK\\
$^4$Institute of Astronomy, Madingley Road, Cambridge, CB3 0HA, UK\\
$^5$Lund Observatory, Division of Astrophysics, Department of Physics, Lund University, Box 43, SE-22100 Lund, Sweden\\
$^6$Princeton University, Department of Astrophysical Sciences, 4 Ivy Lane, Princeton, New Jersey, 08544, United States of America\\
$^7$Research School of Astronomy and Astrophysics, Australian National University, Canberra, ACT~2611, Australia
}
\date{Accepted ---. Received ---; in original form ---}

\pubyear{\date{year}}

\begin{document}
\label{firstpage}
\pagerange{\pageref{firstpage}--\pageref{lastpage}}
\maketitle

\pdfminorversion=5

\begin{abstract}
Motivated by the need for realistic, dynamically self-consistent, evolving galaxy models that avoid the complexity of full, and zoom-in, cosmological simulations, we have developed \nexus, an integral framework to create and evolve synthetic galaxies made of collisionless and gaseous components.
\nexus\ leverages the power of publicly available, tried-and-tested packages: the stellar-dynamics, action-based library \agama;  and the adaptive mesh refinement, N-body/hydrodynamical code \ramses, modified to meet our needs.
In addition, we make use of a proprietary module to account for galaxy formation physics, including gas cooling and heating, star formation, stellar feedback, and chemical enrichment.
\nexus' basic functionality consists in the generation of bespoke initial conditions (ICs) for a diversity of galaxy models, which are advanced in time to simulate the galaxy's evolution.
The fully self-consistent ICs are generated with a distribution-function based approach, as implemented in the galaxy modelling module of \agama\ -- up to now restricted to collisionless components, extended in this work to treat two types of gaseous configurations: hot halos and gas discs. \nexus\ allows constructing equilibrium models with disc gas fractions \mbox{$0~\leq~f_{\rm gas}~\leq~1$}, appropriate to model both low- and high-redshift galaxies.
Similarly, the framework is ideally suited to the study of {\em galactic ecology}, i.e. the dynamical interplay between stars and gas over billions of years.
As a validation and illustration of our framework, we reproduce several isolated galaxy model setups reported in earlier studies, and present a new, `nested bar' galaxy simulation.
Future upgrades of \nexus\ will include magneto-hydrodynamics and highly energetic particle (`cosmic ray') heating.
\end{abstract}

-
\section{Introduction} \label{sec:intro}

The study of the structure and evolution of galaxies can be approached in three different but complementary ways: 1) by direct observation; 2) by theoretical work; 3) by simulation. In the last case, most of the work falls on one of two broad categories: That where the evolution of a significant volume of the Universe and the galaxies within are considered -- so-called {\em cosmological} simulations, and at the other end, that where the evolution of individual galaxies devoid of the boundary conditions (large-scale structure and its gravitational potential, mass and radiation inflows) naturally provided by cosmological simulations is looked at, henceforth referred to as {\em idealised} simulations.

Cosmological simulations have greatly helped advance our understanding of galaxy formation \citep[for an extensive review see][]{naa17a}. The core idea is to evolve gravitationally the inhomogeneities imposed at the start of the simulation on an otherwise uniform matter distribution, which are consistent with the quantum fluctuations in the nascent Universe \citep{zel70a}. Together with the use of a number of `recipes' to mimic otherwise unresolved physical processes believed to be key in shaping galaxies (e.g. star formation and stellar feedback, chemical enrichment, accretion of matter onto black-holes and their feedback, radiative cooling at atomic scales, cosmic ray production, and so forth) this approach allows us to calculate from first principles how galaxies with properties astonishingly similar to real galaxies -- statistically speaking -- emerge across cosmic time \citep[e.g.][]{sch15a,dub16r,gra17h,dol17a,pil18l}.

However remarkable, cosmological simulations also have limitations. The most crucial ones are perhaps: 1) a relatively low resolution (both spatially and in terms of particle sampling), and 2) that they offer no real control over the galaxy properties of interest. Note that while so-called cosmological `zoom-in' simulations \citep[e.g.][]{nuz14a,hop14b,kim14b,wet16a,age21l} do allow for a significant improvement in resolution, they do not mitigate the second issue. For instance, in the study of systems with very specific properties, e.g. Milky-Way (MW) `analogues', compromises have to be made because it is virtually impossible to find systems that satisfy the required constraints (e.g. a galaxy with a dark matter halo mass, a stellar and gas discs masses, and a bulge mass matching those of the MW within their observational uncertainties). This is even more severe if one aims at finding systems that are identical but in one or two aspects, say two MW `twins' that differ only in their stellar mass or their accretion history \citep[but see][]{rot16a,rey18a}. In a nutshell, cosmological simulations allow for a robust statistical analysis of galaxy properties, but are less useful when it comes to study in detail specific systems (e.g. the MW or the Andromeda galaxy, to name a few).

To overcome this problem, there exists an alternative approach that consists in constructing initial conditions for isolated galaxies with very specific properties (e.g. mass, size, structure, etc.), and evolving them in a controlled way. By systematically varying one of the relevant features of the model (e.g. the disc mass) or the way it is evolved (e.g. adiabatic vs. cooling / heating) -- a powerful approach we refer to as\footnote{A differential approach can be applied to cosmological simulations as well \citep[e.g.][]{sch10a} } {\em differential} -- one can isolate the effect of that feature on the overall response of the system \citep[e.g.][]{ath92b,her93b,Wada2001,dim05a,age09a,Hopkins2011,Grisdale2017,Renaud2021,van22a}. It is important to emphasise that these controlled simulations cannot explain how galaxies form, only (at best) how they evolve in
isolation,\footnote{Idealised galaxies can also have companions, i.e. perturbers, but the system as a whole is still isolated.} starting from somewhat {\em ad hoc} initial conditions (hence `idealised'). Needless to say, these three approaches -- cosmological simulations, zoom-in simulations, and controlled simulations of idealised galaxies -- are all complementary to each other.

A requirement to perform controlled simulations is the existence of a framework that allows to create (and evolve) a galaxy model, which accounts for all relevant galaxy components as required by the problem at hand; and most importantly, which is physically self-consistent in the sense that the system's properties (total potential, mass distribution, kinematic structure) are compatible with one another at all times.

A quick review of the literature reveals that there is a wealth of methods (and their publicly accessible implementations in the form of software) available to accomplish the task of setting up an idealised galaxy.\footnote{There is also a significant number of codes to evolve these models; but their discussion is outside the scope of the present paper.} However, those that allow to initialise systems containing both collisionless {\em and} gas components are rare; notable exceptions are MakeNewDisk \citep{spr05c}, the Disc Initial Conditions Environment \citep[\dice; ][]{per14c,per16a}, or \galactics{}+Gas \citep{deg19a}, all of which are in demand \citep[cf.][respectively]{nob24a,de-23a,and22b}, but are by no means prevalent.

Perhaps the scarcity of comprehensive frameworks explains in part the mild aversion of many galactic dynamicists/simulators to consider galaxy models that include a gaseous component. Needless to say, in none but very few cases is this omission justified, even less so when the goal is to interpret observational data of specific systems, for instance, the origin of the \gaia\ `phase spiral' \citep[e.g.][]{lap19a,bla19a,hun21t}, discovered by  \citet{ant18b}. The importance of including gas when studying galaxy structure and galaxy dynamics -- in addition to the obvious reason that stars {\em form} out of gas -- has been unambiguously expressed by \citet{bin08a}, citing a statement attributed to the astronomer Jan Oort: ``The principal features that distinguish lenticular or S0 galaxies from spirals are the low density of cold interstellar gas, the absence of young stars, and the absence of spiral arms. Only a tiny fraction of gas-poor disc galaxies exhibit spiral arms [\ldots] Thus, even though spiral structure is present in the old disc stars, {\em interstellar gas is essential for persistent spiral structure}'' \citep[see also][]{don13a,gho15a}.

Equally startling is the fact that, among the available frameworks to initialise idealised galaxies, only a few follow a distribution function (DF) based approach to fulfil this task, the only approach that yields fully self-consistent results.\footnote{For a discussion, see \citet[][their sec.~5]{vas19a}. }
Indeed, a recurrent issue with simulations of idealised galaxies has been that the initially specified, multi-component system is not in dynamical equilibrium, and thus evolves to a new configuration before long-term stability is achieved; the simulator must then accept a model that is substandard and not what was specified.  This has been a longstanding problem that a DF-based approach, as implemented in
\galactics\ \citep{wid08a}, {\small MKGALAXY} \citep{mcm07a}, or the Action-based Galaxy Modelling Architecture library \citep[\agama;][]{vas19a}, inherently avoids. It is worth stressing that action-based, equilibrium, non-evolving models are relevant in their own right, useful in the study of the observed dynamical and kinematic properties of galaxies \citep[e.g.][]{pif15a,bin23a,gho23b}. But our main focus here is on their use as a starting point to calculate the evolution of idealised galaxies under controlled conditions.

In its original form, \galactics\ only allowed to generate initial conditions (ICs) for collisionless components, appropriate to model systems made of e.g. dark matter and stars. Given the need for more realistic galaxy models that incorporate gas, the library was later extended \citep{deg19a}. The situation is analogous with the \agama\ library, which in its standard form does not allow including responsive gas components. This is a shortcoming our present paper is set out to remedy.

Taking full advantage of the library's machinery, in this work we expand the \agama\ self-consistent modelling (SCM) module to treat gas components in addition to the already included collisionless ones. This is a major innovative step that allows to construct N-body/hydrodynamical models of galaxies that are fully self-consistent from the outset. Crucially, this extension to the \agama\  library, coupled to the adaptive mesh refinement (AMR), N-body/hydrodynamical code \ramses\ \citep{tey02a}, constitutes the backbone of our framework to perform controlled simulations of idealised galaxies, called \nexus.

The main purpose of the present paper is to formally introduce the framework, supplying all relevant details (including those that in earlier studies are only glossed over), as well as to provide a reference for future work. We anticipate that \nexus\ will become useful to the astrophysical community working with these type of simulations. In fact, our group has already made use of the framework in several instances \citep[][]{tep22x,bla23a,bla24a}, and others have followed with a similar approach \citep{ann24a}.\\

The structure of this paper is as follows: Sec.~\ref{sec:gasconf} describes some of the most widely used methods to create equilibrium galactic gas configurations. Sec.~\ref{sec:imp} describes our implementation of a subset of these methods within the \agama\ library. In Sec.~\ref{sec:galevol} we briefly describe the modifications to the \ramses\ code required by our framework, including our adopted implementation of galaxy-formation physics. In Sec.~\ref{sec:val} we validate our method with a number of simple test cases, and present a case of astrophysical interest in Sec.~\ref{sec:s2b}. We conclude with a reflection about the importance of controlled simulations of idealised galaxies in Sec.~\ref{sec:conc}.

\section{Idealised, gaseous galactic components} \label{sec:gasconf}

In what follows, we present an overview of some of the most common approaches found in the literature to construct equilibrium galactic gas configurations \citep[cf.][]{rec14a}. Specifically, we focus on models of galactic hot halos (or `coronae') and galactic gas discs. Throughout, we limit our scope to ideal, mono-atomic gases. We refrain from discussing the setup of collisionless components with \agama, which has been discussed at length elsewhere \citep[cf.][]{vas19a,bla21e,tep21v,tep22x,bla23a,bla24a}.
 
\subsection{Galactic coronae} \label{sec:gcorona}

Originally predicted by \citet{spi56a} as the confining medium around neutral-hydrogen (\HI) clouds in the Galactic halo, and later by early theories of galaxy formation \citep{bin77c,whi78a}, hot gas halos (or galactic `coronae') have now been firmly established in a handful of systems, including the Milky Way \citep{mil16a}, Andromeda \citep{leh17a}, and most recently the Magellanic Clouds \citep{kri22a}. This suggests that likely every spiral galaxy in the Universe is embedded within a corona of hot gas \citep[for a comprehensive review, see][]{bre18a}.

The importance of galactic coronae cannot be overstated. They are believed to constitute significant gas reservoirs from which galaxies may obtain sufficient gas supply to maintain relatively high star formation over cosmologically relevant time scales \citep[e.g.][]{mar10b,mos11a,hwa13a,gro18a}. In addition, they are a dynamically important component, which affects the accretion of gas onto the star-forming disc delivered through cosmological filaments \citep{ste24a} or via tidal disruption of satellites \citep{mas05a,tep18b,tep19a}. Also, galactic coronae may be key in providing a solution to the `missing baryons' at low redshift \citep{fuk06a}.
Finally, they provide external pressure onto the dense interstellar medium (ISM), leading to a more complex disc-halo interaction \citep[e.g.][]{arm16a}.

In short, galactic coronae are a physical reality, and a necessary ingredient in any realistic galaxy model, and yet they are generally ignored. This deficiency of many idealised galaxy models is one of the many we intend to address with our framework.

\subsubsection{Halo density structure and internal energy} \label{sec:ghdens}

Galactic coronae are often --  out of convenience -- assumed to be spherical, isothermal gas configurations at a temperature $T$ in hydrostatic equilibrium (HSE) with the total galactic potential $\Phi$ \citep[][]{sut07a}. In this case, their density structure as a function of the spherical radius $r = \sqrt{R^2 + z^2}$ -- with $R$ and $z$ the cylindrical coordinates -- is described by
\begin{equation} \label{eq:hhse}
	\rho_{\rm gh}(r) = \rho_0~\exp \left\{ \Phi(R,z) / c^2_s \right \} \, .
\end{equation}
Here, $\rho_0$ is the central density, $c^2_s = k_b T / \mu~m_p$ is the isothermal sound speed, and $\Phi \leq 0$.

However, such models are far from realistic, as suggested both by observations \citep[][]{hod18a} and cosmological simulations \citep[][]{opp18a}. Yet, they are very useful for the purposes of validating a method aimed at creating and evolving such configurations, given that the analytic solution is known, with the potential caveat that HSE is notoriously difficult to maintain in fluid dynamics (grid) codes \citep[][]{kap16a,kra19a,can22a}.

Less restrictive -- and at the same time more realistic -- models correspond to spherical configurations described by a density $\rho_{\rm gh}(r)$ in HSE with the total potential, but not necessarily isothermal. For these, the temperature profile can be obtained from the \citet{jea15a} equation for a spherically symmetric system,
\begin{equation} \label{eq:sigma}
	\sigma^2_r(r) = \frac{ 1 }{ \rho_{\rm gh} } \int_r^\infty \rho_{\rm gh} \frac{ \partial \Phi }{ \partial r'} dr' \, ,
\end{equation}
where the (no longer constant) sound speed is set identical to the macroscopic velocity dispersion, i.e. \mbox{$c^2_s \equiv \sigma^2_r$} \citep[cf.][]{mas05a}, implying a (spherically symmetric) temperature profile
\begin{equation} \label{eq:thse}
	T_{\rm HSE}(r) = \frac{ \mu~m_p }{ k_b } \sigma^2_r(r) \, .
\end{equation}
%
Yet more realistic models correspond to partially pressure-supported, {\em spinning} coronae \citep[][]{bar06a,pez17a,sor18a} in vertical hydrostatic equilibrium (vHSE; $v_z \equiv 0$), in which the temperature at any given radius is dictated by the balance between the hydrostatic pressure support and the azimuthal velocity $v_\phi$,
\begin{equation} \label{eq:trot}
	T_{\rm rot}(r) = T_{\rm HSE}(r) - \frac{1}{2}(1 - \gamma) \frac{ \mu~m_p }{ k_b } v^2_\phi \, . 
\end{equation}
Here, $\gamma $ is the adiabatic index and is equal to 5/3 for a mono-atomic gas. This factor appears because the balance between pressure and rotational support is obtained in terms of the specific internal energy of the gas,
$$e = c^2_s / (1 - \gamma) \, .$$
Thus, the model is fully specified once $\rho$ and $v_\phi$ are.

At galactic scales, this last type of models is certainly the most relevant, and for this reason they form the basis of the corona models included in our framework (see Sec.~\ref{sec:gcorona_agama}). \\

It is worth noting that the approach outlined above always yields spherically symmetric configurations. However, spinning coronae will flatten, adopting an oblate or even a toroidal geometry \citep[e.g.][]{tep18b}. Such models can be setup from the outset \citep[e.g.][]{pez17a,sor18a}, but we defer these cases to future versions of \nexus, where we will also consider triaxial (dark matter) halos \citep[e.g][]{ath13l}.

\subsubsection{Halo velocity structure} \label{sec:ghvel}

In the case of spinning coronae in vHSE, only the $v_x$- and $v_y$-velocity components need to be specified, which in turn requires knowledge of the azimuthal velocity $v_\phi$. The latter is however not constrained in any way by the \citet{jea15a} equations, and can thus be freely chosen. 

Different approaches have been adopted in the past. A widely used approach, pioneered by \citet[][]{str00a}, relies on setting $v_\phi = e v_c$, where $0 < e < 1$ and $v^2_c = R ~\partial \Phi / \partial R$ is the total circular speed of the system \citep[e.g.][]{bar23a}.

A different, more physically motivated, approach involves assuming a self-consistent centrifugal support where the velocity is obtained by solving for $v_\phi$, given a density $\rho$ and an equation of state $P(\rho)$ (e.g. \citealt[][]{der99a}; see also Sec.~\ref{sec:ggdisc}).

Yet another common approach, motivated from cosmology, is based on the following idea. Tidal torque theory suggests that a collapsing dark matter (DM) halo, and the baryons within, acquire angular momentum ($J$) as a result of a misaligned (i.e. not perfectly radial) infall \citep{hoy49a,whi78a,efs79a,fal80a}. The specific angular momentum $j \equiv J / M$ of the virialised system of total mass $M$ (DM halo and baryons) can be indirectly parameterised by the `oblatness' parameter \citep{pee69b}, more popularly known as `spin' parameter \citep[][]{mo98a},
\begin{equation} \label{eq:spin}
	\lambda^2 \equiv  \frac{ j^{\,2} ~\vert E \vert }{ G^2 ~M^{3} } \, ,
\end{equation}
where $E$, is the sum of potential, kinetic and thermal energies.

The core idea is that the DM halo and the baryons within have a spin determined by either $\lambda$ or $j$, which is in turn set by the fundamental properties ($M$, $E$) of the collapsing system. For instance, cosmological, DM-only simulations of structure formation suggest that $j \propto M^s$ with $s = 1.3 \pm 0.3$, where $M$ is the total mass enclosed within $r$ \citep{bul01b}. It can be shown that such a behaviour derives from an exponential angular momentum distribution, $\psi(l) \propto (M / l)~e^{-l/l_0}$ \citep[][]{sha12b}.

Estimates for $\lambda$ are also obtained from these simulations, which indicate that DM halos have spin parameters that roughly follow a log-normal distribution with its peak at approximately $0.035 \pm 0.005$ and width $0.5 \pm 0.03$ \citep{bul01b}. Full hydro-cosmological simulations indicate that the spin of (hot) gas accreting onto a DM halo displays a similar behaviour with mass, but with a higher normalisation, up to factors~3 compared to the DM, i.e.~$\langle \lambda_{gas} \rangle \approx 0.12$  \citep{pic11a}. The latter parameter values are adopted as the default values within our framework.

\subsection{Galactic gas discs} \label{sec:ggdisc}

Settled, galactic gas discs are generally observed to follow an exponentially declining surface density with a characteristic scale length $R_d$ \citep[][]{ler09a,kal09a},
\begin{equation} \label{eq:sig1}
	\Sigma(R) = \Sigma_0 \exp\{ -R / R_d\} \, .
\end{equation}
Thus, many of the approaches in the literature used to set up isolated, galactic gas discs assume such a profile; the vast majority of them follow the method pioneered by \citet*{spr05c}. In essence, their approach consists in solving the steady-state momentum conservation equation for a perfectly axisymmetric ($\partial / \partial \phi \equiv 0$), rotationally supported ($v_R \equiv 0$) disc in vertical hydrostatic equilibrium ($v_z \equiv 0$),
\begin{align}
	\frac{ 1 }{ \rho } \frac{ \partial P }{ \partial R } + \frac{ \partial \Phi }{ \partial R }&  = \frac{ v_\phi^2 }{ R }, \label{eq:radbal} \\
	\frac{ 1 }{ \rho } \frac{ \partial P }{ \partial z } + \frac{ \partial \Phi }{ \partial z }&  = 0 \, ,\label{eq:zbal} 
\end{align} 
for the density $\rho(R,z)$, given a total galactic potential, $\Phi$, and a specified equation of state $P(\rho)$, and demanding that the density satisfy Eq.~(\ref{eq:sig0}; see below).

Here, we adopt an alternative approach, introduced by \citet{wan10a}. These authors developed a simple iterative, computationally efficient method to arrive at a solution for the special case of an {\em isothermal} gas disc, characterised by the equation of state \mbox{$P(\rho)~=~c_s^2~\rho$} with a constant sound speed $c_\mathrm{s}$, a fixed density normalisation $\Sigma_0$, and fixed radial scalelength $R_d$ \citep[see also][]{rod11y}.

Although their method is only strictly valid for isothermal configurations, which are of interest in their own right as examples of perfect equilibrium systems used for validation purposes (see Sec.~\ref{sec:val}), it also serves well to set up a galactic gas disc intended {\em as a starting point} of a more realistic system that is allowed to undergo cooling and heating, and to form stars. Such a system will not maintain its initial thermal state nor its initial density structure, but it will instead evolve into a different, likely quasi-stable configuration regulated by interplay between energy injection and dissipation. Importantly, the gas will have no memory of its (rather ad hoc) initial thermal and structural configuration after a few hundred million years \citep[cf.][their sec. 4]{bla24a}.\\

In what follows, we briefly outline the method put forward by \citet{wan10a}. In Sec.~\ref{sec:ggdisc_agama}, we describe how the algorithm is adapted to our needs.

\subsubsection{Disc density structure} \label{sec:gddens}

Given that $c_s$ is a constant by assumption, Eq.~\eqref{eq:zbal} can be readily integrated to yield the gas volume density,
\begin{equation} \label{eq:rhogd}
	\rho(R,z) = \rho_0(R) ~\rho_z(R,z) \, ,
\end{equation}
where
\begin{equation} \label{eq:rhogdz}
	\rho_z(R,z) \equiv \exp \left\{ -\Phi_z / c^2_s \right\} \, ,
\end{equation}
and 
\begin{equation} \label{eq:phiz}
	\Phi_z(R,z) \equiv \Phi(R,z) - \Phi(R,0) \, .
\end{equation}
Integrating Eq.~\eqref{eq:rhogd} over $z$ yields the gas surface density,
\begin{equation} \label{eq:sig0}
	\Sigma(R) = \int_{-\infty}^{+\infty} \!\!\!\!\! \rho(R,z) ~dz = 2 ~\rho_0(R) \!\! \int_0^{+\infty} \!\!\!\!\! \rho_z(R,z) ~dz \, ,
\end{equation}
where in the last step it is assumed that the potential, and thus the volume density, is symmetric with respect to the mid-plane.

Mathematically, the factor $\rho_0$ in Eq.~\eqref{eq:rhogd} is a constant of integration; physically, it corresponds to the mid-plane gas density. Using Eq.~\eqref{eq:sig0}, it can be expressed as
\begin{equation} \label{eq:rhogd0}
	\rho_0(R) =  \frac{ 1 }{ 2 }\Sigma(R) \left[ \int_0^{+\infty} \rho_z(R,z) ~dz \right]^{-1} \, .
\end{equation}
Modelling the structure of a strict isothermal gas disc thus reduces to the problem of finding $\rho_0(R)$ and $\rho(R,z)$ for a given $\Sigma(R)$, and a corresponding $\Phi_g$ via Poisson's equation,
\begin{equation} \label{eq:poig}
	\nabla^2 \Phi_{g} = 4 \pi G \, \rho \, ,
\end{equation}
such that Eqs.~(\ref{eq:rhogd} -- \ref{eq:rhogd0}) are all self-consistent. Here, $\Phi_g$ is the contribution of the gas disc to the total potential of the system,
\begin{equation} \label{eq:tpot}
	\Phi = \Phi_g + \sum_i \Phi_i \, ,
\end{equation}
where $\Phi_i$ are any other galaxy components (e.g.~DM halo, stellar bulge, stellar disc).

\citet{wan10a} provide an iterative, efficient algorithm to accomplish this task, which can be summarised as follows. First, fix $\Sigma(R)$ and the contributions $\Phi_i$ to the total potential $\Phi$, and choose an initial guess for $\rho_0(R)$. Then:
\begin{enumerate} \label{par:witer}
	\item[(1)] express $\rho$ using $\rho_0$ and Eqs.~\eqref{eq:rhogd} - \eqref{eq:phiz}, then
	\item[(2)] solve\footnote{ \citet{wan10a} use an approximate form of Poisson's equation, valid only for thin discs, \mbox{$d^2 \Phi_{g} / dz^2 = 4 \pi G \rho(R,z)$}. However, this approximation is not necessary within our framework (see Sec.~\ref{sec:ggdisc_agama})} for $\Phi_{g}$ via Eq.~\eqref{eq:poig}, then
	\item[(3)] calculate a new $\rho_0$ via Eq.~\eqref{eq:rhogd0}.
	\item[$\to$] Go to step (1) and repeat until convergence.
\end{enumerate}
In this context, `convergence' can be measured, for example, by tracking the relative change in the value of the central density $\rho_0(R=0)$ and demanding it falls below some prescribed threshold, say 1 percent. In such a case, the iterative process converges in a few iterations, provided a suitable choice is made for the initial guess $\rho_0(R)$. For instance, \citet{wan10a} recommend setting $\rho_0(R) \propto \Sigma(R)$, which works remarkably well.\\

As anticipated by \cite{deg19a}, the above algorithm can be seamlessly integrated into \agama's self-consistent modelling iterative procedure, which we describe below in Sec.~\ref{sec:imp}.
 
\subsubsection{Disc velocity structure} \label{sec:gdvel}

Once the density structure of the disc and the total potential of the self-consistent model are known, $v_\phi$ is fully determined by
\begin{equation} \label{eq:vphig}
	v^2_\phi = \frac{ R }{ \rho } \frac{ \partial P }{ \partial R} + v^2_c \, ,
\end{equation}
which follows from Eq.~\eqref{eq:radbal} with \mbox{$v^2_c \equiv R~\partial \Phi / \partial R$}.
Since $P$ (and $\rho$) decreases with radius for an exponential disc, the azimuthal gas velocity is the result of the balance between the pressure gradient and the centrifugal force induced by the total galactic potential $\Phi$. 

In general, the gas rotation velocity is dependent both on the cylindrical radius, $R$, and the height from the mid-plane, $z$ \citep[][]{bar06a}, but it depends only on $R$ for barotropic (i.e.~$P = P(\rho)$) configurations \citep[{\em Poincaré-Wavre theorem}; q.v.][]{leb67a}, which include isothermal configurations as a special case. Thus, for all our intents and purposes, Eq.~\eqref{eq:vphig} defines the rotation velocity for any $z$, i.e.~\mbox{ $v_\phi(R,z) \equiv v_\phi(R)$ }.

\section{Implementation in \agama} \label{sec:imp}

In the following, we describe our implementation in \agama\ of the approaches discussed above to construct equilibrium models of hot halos and gas discs. We start by providing a brief overview of the relevant theory underlying the DF-based approach to set up initial conditions for isolated galaxy models, with an emphasis on the iterative nature of the problem, and explain how we exploit it for our needs. More details can be found in \citet[][see also \citealt{tep21v}, their sec.~3]{vas19a}.

\subsection{Theoretical background}

The evolution of a collisionless N-body system with a `sufficiently large' number of particles is governed by the collisionless Boltzmann equation and its solution, the `one-particle' distribution function, $f(\bm{r},\bm{v},t)$. In a steady state, \mbox{$f(\bm{r},\bm{v},t') \equiv f(\bm{r},\bm{v},t=0)$}, for any time $t' > 0$, implying that $f$ is a function of the integrals of motion only \citep{jea15a}.

A convenient set of integrals of motion are the action variables, which in the case of axisymmetric potentials, are the radial action $J_R$, the vertical action $J_z$, and the azimuthal action $J_{\phi}$ (equivalent to the $z$-component of the angular momentum vector). In this case,\footnote{ A bold $\bm{J}$, denoting a set of actions, should not be confused with a regular $J$, denoting total angular momentum. } \mbox{$f = f(\bm{J}) \equiv f(J_R,J_z,J_\phi)$}. This implies the existence of a mapping \mbox{$(\bm{r},\bm{v}) \mapsto \bm{J}$} that depends on the total potential of the system $\Phi$, succinctly expressed as\footnote{The mapping also includes the three angles variables, which we ignore henceforth because the integral over these variables simply yields a factor $(2 \pi)^3$ which is absorbed into the normalisation of the DF.} $\bm{J}\left[\bm{r},\bm{v} ~\vert ~\Phi \right]$.

The distribution function provides a statistical description of the dynamical state of the system, and other properties can be derived from it. For example, for an appropriate normalisation, the mass density $\rho_c$ of a component (e.g. stellar disc) described by a DF $f_c$ is given by
\begin{equation} \label{eq:dens}
    \rho_c(\bm{r}) = \int \!\! f_c \left\{ \bm{J}\left[ \bm{r},\bm{v}~\vert ~\Phi \right] \right\} ~d\bm{v} \,.
\end{equation}
%

\subsection{Self-consistent models} \label{sec:scm}
 
In \agama, a fully self-consistent, multi-component galaxy model is defined by:
\begin{itemize}
    \item The density of each component, $\rho_c(\bm{r})$, calculated from its corresponding DF ($f_c$)
    \item The DF of each component, expressed in terms of actions, $f_c \left\{ \bm{J}\left[ \bm{r},\bm{v}~\vert ~\Phi \right] \right\}$, dependent on the potential $\Phi$.
    \item The total potential of the system, $\Phi$, determined via the superposition of all $\rho = \sum \rho_c$ via Poisson's equation $\nabla^2 \Phi = 4 \pi G \rho$.
    \item The mapping $\bm{J}\left[\bm{r},\bm{v} ~\vert ~\Phi \right]$, for which \agama\ makes use of the so-called `St\"ackel fudge' \citep{bin12u}, appropriate for axisymmetric configurations.
\end{itemize}
The relationship between these model ingredients imply a coupled system of non-linear equations that must be solved iteratively. To solve it, the self-consistent modelling (SCM) module within the \agama\ library adopts the following algorithm:
\begin{enumerate}\label{par:aiter}
    \item[(1)] Specify the initial (target) density profile
    $\rho_c$ (or alternatively, the corresponding potential), of each galaxy component, then
    \item[(2)] specify the DF, $f_c(\bm{J})$, for each component; $f_c$ remains fixed during subsequent iterations, then
    \item[(3)] calculate the potential of the system via $\nabla^2 \Phi = 4 \pi \sum_c \rho_c$, then
    \item[(4)] determine the mapping $\bm{J}\left[\bm{r},\bm{v} ~\vert ~\Phi \right]$ and recompute the new density of each component $\rho'_c$ via Eq.~\eqref{eq:dens}.
    \item[$\to$] Go to step (3) and repeat until `convergence' (see below).
\end{enumerate}
Convergence during the iterative process is guaranteed by the adiabatic invariant nature of actions, together with the use of $f(\bm{J})$. It can be assessed, for example, by the tracking the maximum relative change across iterations of the total potential at the origin, which ideally should be of the order of a percent or less. 

Once a converged model has been attained, an N-body representation of the system is created by drawing a specified number $N_c$ of random samples (point-like masses or `particles' of equal mass) for each component directly from phase space $(\bm{r}, \bm{v})$ by evaluating $f(\bm{J})$ using the mapping $\bm{J}\left[\bm{r},\bm{v} ~\vert ~\Phi \right]$ together with the final, self-consistent potential. To this end, \agama\ implements an efficient multidimensional sampling-rejection algorithm. Thus, the construction of an N-body model with \agama\ yields a {\em fully self-consistent} \mbox{`potential-density-velocity distribution'} triplet -- a necessary requirement on any method to initialise an idealised galaxy model (cf. Sec.~\ref{sec:intro}). It is worth stressing that self-consistency does not imply a {\em stable} dynamical equilibrium, e.g. a self-consistent model can be bar-unstable from the outset.

\subsection{Galactic coronae} \label{sec:gcorona_agama}

In its standard form, \agama\ readily allows setting up hot a halo as a fully valid, DF-based component, starting from a target density profile. In essence, this is the same approach used to set up a generic `spheroidal' component, say, a DM halo or a classical stellar bulge. The difference is that, once the model has converged (in the sense of Sec.~\ref{sec:scm}), a hot halo component requires some post-processing in order to arrive at a self-consistent gas configuration, as described below.

\subsubsection{Initial density profile and distribution function}

We focus our attention on a partially pressure-supported, spinning hot halo embedded in a DM host halo. We restrict our discussion to galactic DM halos that follow a Navarro-Frenk-White (NFW) profile \citep{nav97a}, and assume for the sake of simplicity that an embedded hot halo follows the same profile (but possibly with a different scaling; see Sec.~\ref{sec:hothalo}). In doing so, we follow on the footsteps of a long list of seminal studies  \citep[e.g.][see also \citealt{aum10a,mos11a,tey13a,cla19a}]{mo98a,asc03a,mas05a,kau06a}. It is worth emphasising that this assumption is adopted out of mere convenience, and that many other choices are possible. Indeed, the SCM module of the \agama\ library provides a large collection of physically motivated, spheroidal profiles, any of which can be adopted for the DM halo or the hot halo when creating a galaxy model within our framework. The choice of a {\em specific}, in this case NFW, profile (as well as a particular kinematic and thermal structure) for the hot halo is simply intended for validation purposes, but the model thus constructed also serves well as a reasonable starting point for a realistic galaxy model (see argumentation before Sec.~\ref{sec:gddens}).

Now, quite generally, the initial density of the hot halo is described by the following function,\footnote{The spheroidal profile in \agama\ has more free parameters than included in Eq.\eqref{eq:sph}, but we list only those that are relevant to our modelling; the omitted parameters all take their \agama\ default values. We refer the reader to the \agama\ documentation \citep{vas18a} for details. \label{fon:doc} }
\begin{equation} \label{eq:sph}
    \rho_s(r) = \rho_0 \left( \frac{r}{r_s} \right)^{-\gamma} \left[ 1 + \left( \frac{r}{r_s} \right) \right]^{\gamma - \beta} \times \exp{\left[ - \left( \frac{r}{r_c} \right)^2 \right]} \, ,
\end{equation}
which describes a double power-law, spheroidal profile with a taper. Here, $r$ is the spherical radius; the other parameters have the following meaning:
\begin{itemize}
    \item $\rho_0$ := density normalisation
    \item $r_s$ := scale radius
    \item $r_c$ := outer cut-off radius
    \item $\beta$ := outer power-law slope
    \item $\gamma$ := inner power-law slope
\end{itemize}
We note that this density profile is available as part of the  standard \agama\ Potential-Density Factory, and it is flexible enough to accommodate a significant number of relevant profiles (we refer the reader to the \href{http://agama.software}{\agama\ documentation} for more details).

To arrive at a density configuration that closely (or exactly, if the hot halo is isolated) resembles a NFW profile, one must set $\beta \equiv 3$ and $\gamma \equiv 1$. The remainder of the parameters determine the mass, concentration, and extension of the halo, and they should be chosen according to the intended application.

During the iterative process (Sec.~\ref{sec:scm}), we assign the halo a DF of the type `Quasi-Spherical', well-suited to model spheroidal components. However, instead of specifying the analytic form of the DF (which is possible with \agama), the DF is constructed directly from the density distribution \eqref{eq:sph} using either the Eddington inversion formula or its anisotropic generalisation \citep{cud91a}, which has the advantage of reducing the number of free parameters.

Once the mass distribution of the hot halo is fixed and converged, all that remains to be calculated is the velocity structure and the internal energy, in that order (see Eq.~\ref{eq:trot}), which is done in post-processing, using external tools that add functionality to the standard \agama\ library

\subsubsection{Post-processing} \label{sec:ppgh}

Post-processing is accomplished with help of the \pynbody\ package \citep{pon13a} coupled to \agama's SCM output. During this procedure, the particle constituents of the hot halo are assigned a temperature $T$ (or equivalently, an internal energy $e$). Furthermore, depending on the intended application of the model, the gas may be required to have a specific metal mass fraction $Z_{met}$ (or equivalently, a specific elemental abundance distribution), e.g. if the gas is allowed to heat, to cool (radiatively), and to form stars (cf. Sec.~\ref{sec:galevol}).

In the case of a {\em uniform} metal mass fraction (or specific elemental abundance) -- which we focus here on,\footnote{Whether a uniform chemical pattern is appropriate for the intended application is a different matter, whose discussion is beyond the scope of this paper.} the assignment is trivial as it amounts to create additional particle properties (i.e. effectively new arrays) with the desired values, and it is not discussed further. In contrast, assigning a temperature profile requires some care as it entails calculating first the velocity structure, and then re-assigning the macroscopic (i.e. particle) velocities obtained via the self-consistent DF-based approach accordingly. It is worth stressing that, since the density distribution is not altered in any way during post-processing, the system retains its full dynamical consistency, provided the relation between velocities and temperature is self-consistent as well (Eq.~\ref{eq:trot}).

As outlined in Sec.~\ref{sec:ghvel}, the velocity structure is calculated assuming that the specific angular momentum of the gas, $j_h$, follows that of the dark matter \citep[c.f.][]{kau06a}. Specifically, we adopt $s = 1$, which implies
\begin{equation} \label{eq:vphigh}
	v_\phi(r) = \frac{ j_h(r) }{ r } = g_0 \frac{ M_h(<r) }{ r } \, ,
\end{equation}
where $g_0$ is an appropriate normalisation constant to be determined.

To this end, first we need to calculate the total angular momentum, $J$, of the system. Because we are assuming the gas follows the same profile as the DM, we use the {\em total} (DM + gas) mass $M$ to calculate the virial radius $r_{\rm vir}$; we fix the value of the spin parameter $\lambda$, and calculate the energy via \citep{mo98a}
$$E_{\textnormal{\sc dm}} = - \frac{ 1 }{ 2 } \frac{G M^2_{\textnormal{\sc dm}} }{ R_{\textnormal{\sc dm}} } f_c \, ,$$
where
$$f_c = \frac{ c }{ 2 } \left[ 1 / x^2 - (2 \ln x) / x \right] ~\left[ c/x^{\phantom{2}} - \ln x  \right]^{-2} \,,$$
$x = 1 + c$, and $c = r_{\rm vir} / r_s$. We note that these expressions are strictly valid only for NFW halos. If the DM (or hot) halo is required to follow a different profile, then different formulae are necessary. The total angular momentum $J$ is then calculated from Eq.~\eqref{eq:spin}, using the values of $M$, $r_{\rm vir}$, and $E$ just obtained.

Now let the mass of the DM halo and of the hot halo be a fraction $m_{\textnormal{\sc dm}}$ and $m_h$, respectively, of the total mass $M$, i.e. \mbox{$M_{\textnormal{\sc dm}} = m_{\textnormal{\sc dm}} M$} and \mbox{$M_{h} = m_h M$}, such that \mbox{$m_{\textnormal{\sc dm}} + m_h  = 1$}; and apportion the total angular momentum $J$  in the same way, i.e. \mbox{$J_{\textnormal{\sc dm}} = m_{\textnormal{\sc dm}} J$} and \mbox{$J_{h} = m_h J$}. The value of the constant $g_0$ of interest here\footnote{The same procedure can be applied to the DM halo, yielding a {\em different} value for $g_0$.} is such that \mbox{$g_0 J_h \equiv m_h J$}. To calculate its value, we proceed as follows \citep[cf. \dice; ][]{per14c}: We assign temporarily the halo particles a velocity according to Eq.~\eqref{eq:vphigh} setting $g_0 \equiv 1$. Then, we use the $v_\phi$ thus obtained to calculate the space velocity $\vec{v}$ of each gas particle of mass $m_g$ located at $\vec{r}$, and use these to calculate the magnitude of the angular momentum via \mbox{$J_h = m_g ~\vert ~\vec{r} \times \vec{v} ~\vert$}, and thus finally $g_0 = m_h (J / J_h)$.

We are now in the position to calculate the actual $v_\phi$ according to Eq.~\eqref{eq:vphigh}, and to assign each gas particle a macro-velocity with components
\begin{equation} \label{eq:vel1}
	v_x = - v_\phi ~\sin \theta \quad ; \quad v_y = v_\phi ~\cos \theta  \quad ; \quad \left[ v_z \equiv 0 \right] \, .
\end{equation}
where $\theta$ is the particle's polar angle and is given by $\tan \theta = y / x$, with $(x,y)$ the particle's Cartesian in-plane coordinates. Our choice of sign yields a counter-clockwise rotating halo if observed downwards along the $z$-axis. The last of these equations in square brackets is imposed by the condition of vertical hydrostatic equilibrium (vHSE), and is only used if such a condition is to be enforced. Otherwise, the vertical velocity must be determined from other physical constraints. Note that the definitions of $v_x$ and $v_y$ ensure that initially there is no radial motion ($v_R \equiv 0$). 

Finally, the temperature profile (internal energy) is then calculated via Eq.~\eqref{eq:trot}, and assigned to each gas particle as an additional property depending on its position ($r$).

\subsection{Gas discs} \label{sec:ggdisc_agama}
 
In contrast to hot halos, a rotationally supported gas disc in vertical hydrostatic equilibrium must be included in the SCM iterative procedure as a `static' (as opposed to DF-based) component. Based on the discussion presented in Sec.~\ref{sec:gddens}, a reasonable {\em initial} density profile for the gas disc is the radially-exponential, vertically-isothermal model (also available as one of the built-in profiles in \agama, in addition to others):
\begin{equation} \label{eq:disc}
    \rho_d(R,z) = \frac{\Sigma_0}{4 | z_d |} \exp{ \left[ - \frac{R}{R_d} \right]} \times \sech^2{\left| \frac{z}{2 z_d} \right|} \, ,
\end{equation}
where $R$ is the cylindrical radius, $z$ is the vertical distance from the plane, and the other parameters have the following meaning:
\begin{itemize}
    \item $\Sigma_0$ := surface density normalisation
    \item $R_d$ := radial scalelength
    \item $z_d$ := vertical scaleheight
\end{itemize}

Once the initial density profile of the gas disc has been specified and its temperature $T$ (or equivalently $c^2_s$) has been fixed, the procedure to set up a full model looks as follows (cf. \ref{par:aiter} and \ref{par:witer}):
\begin{enumerate}
    \item[(1)] Specify the initial density profile $\rho_c$ of each galaxy component, both collisionless and gaseous, then\\[-8pt]
    \item[(2)] specify the DF, $f_c(\bm{J})$, for each component; $f_c$ remains fixed during subsequent iterations, then\\[-8pt]
    \item[(3)] calculate the potential of the system via $\nabla^2 \Phi = 4 \pi \sum_c \rho_c$, then\\[-8pt]
    \item[(4)] determine the mapping $\bm{J}\left[\bm{r},\bm{v} ~\vert ~\Phi \right]$ and recompute density of each {\em DF-based} component $\rho'_c$ via Eq.~\eqref{eq:dens}, then\\[-8pt]
    \item[(5)] recompute the density $\rho'_c$ of the {\em static} gas disc (Eq.~\ref{eq:rhogd}) by recalculating \\[-15pt]
    	\begin{enumerate}
		\item the vertical density $\rho_z(R,z)$ via Eqs.~\eqref{eq:rhogdz} and \eqref{eq:phiz},
		\item the mid-plane density $\rho_0(R)$ via Eq.~\eqref{eq:rhogd0}. \\[-12pt]
	\end{enumerate}
    \item[$\to$] Go to step (4) and repeat until convergence.\footnote{Note that steps (4) and (5) are conceptually similar, differing only in the way the density is recomputed (either from an action-based DF for stars or from the hydrostatic equilibrium for gas).} 
\end{enumerate}
In its standard form, \agama\ takes care of steps (3)-(4), and provides a number of options for the user to complete steps (1) and (2). However, steps (5a-b) require adding functionality to the library. Specifically, functions to evaluate $\rho_0(R)$ and $\rho_z(R,z)$ must be provided. In our implementation, we integrate the denominator in Eq.~\eqref{eq:rhogd0} using a simple trapezoidal integration, which considerably speeds up the calculations and thus the iterative process.

\subsubsection{Post-processing} \label{sec:ppgd}

As is the case with hot halos, gas discs  require some post-processing upon convergence of the \agama\ model. In brief, one needs to calculate $v_\phi$ (Eq.~\ref{eq:vphig}), and assign each of the gas particles a macroscopic velocity according to Eq.~\eqref{eq:vel1}, and set $v_z \equiv 0$ if the disc is to be in vHSE. Note that, as with the halo, our choice of sign yields a counter-clockwise rotating disc.

In addition, it is necessary to impose a temperature (or internal energy) as well as a global fraction of heavy elements (or individual elemental abundances) onto the gas particles. Since we are dealing with isothermal discs, the former task is simply fulfilled by attaching the (constant) temperature $T$ -- corresponding to $c^2_s$ used to calculate $\rho_z$ (Eq.~\ref{eq:rhogdz}) -- to each particle as an additional property (i.e. effectively an additional array). The latter task will depend on the particular goal of the model, and may be as simple as assigning a constant metallicity to each particle, or as complex as assigning each particle a specific elemental abundance based on its position within the synthetic galaxy, e.g. a radial metallicity gradient.\\

Collisionless components created with \agama\ do not require any post-processing. Nonetheless, we have found it advantageous for analysis purposes to assign each constituent component (e.g. DM halo, bulge, stellar disc, etc.) a unique tag `{\sc comp\_id}' (effectively an {\em odd} integer number)\footnote{ We choose e.g. {\sc comp\_id[dm]}=1, {\sc comp\_id[bulge]}=3,  {\sc comp\_id[disc]}=5, and so on. } as well as to assign each particle a unique number `{\sc part\_id}', that serves both to distinguish the particle from the rest {\em and} to identify the particle as belonging to a particular component. For each component identified by {\sc comp\_id}, the latter is accomplished by looping over the particles numbered $i = 0, 1, 2, \ldots$ and assigning particle $i$ the following
$$
	\textnormal{\sc part\_id} = \left( 2~\textnormal{\sc comp\_id}-1 \right) + 2i~\textnormal{\small MaxCompNumber} \,,
$$
where `{\small MaxCompNumber}' is the largest number of components expected in any reasonable galaxy model. For backwards compatibility, we have set \mbox{{\small MaxCompNumber} = 11}. Note that each {\sc part\_id} is unique across components within a given galaxy model, and it is always an odd, integer number.

The particles belonging to a component identified by  {\sc comp\_id} can be easily retrieved by selecting those for which the condition
$$
	\textnormal{ mod({\sc part\_id}, 2{\small MaxCompNumber}) }~\stackrel{!}{=}~\textnormal{\sc comp\_id}
$$
is satisfied, which is easily accomplished with the help of \pynbody.

\section{Galaxy evolution with Ramses} \label{sec:galevol}

The basic workflow of \nexus\ consists in creating initial conditions for a desired galaxy model, and evolve them with a modified (`patched') version of \ramses. The simulation output is processed (analysed/visualised) with a custom Python package, Ramses Analysis and Visualisation Environment ({\small RAVE}),\footnote{{\small RAVE} will soon be made publicly available.} built around \pynbody\ and the Python interface to the {\tt Tcl/Tk} Graphical User Interface (GUI) toolkit, {\sc Tkinter}.\footnote{Written by Steen Lumholt, Guido van Rossum, and Fredrik Lundh (\url{https://docs.python.org/3/library/tkinter.html}). }

Our relevant modifications to the standard \ramses\ code go into the modules to read in and process the ICs; and the modules to treat the evolution of the gas, including cooling, heating, star-formation and stellar feedback.

The initial conditions are stored in Gadget-2 (binary) format \citep{spr05b} using the Universal N-body Snapshot Input/Output ({\small UNSIO}) library.\footnote{Written by Jean-Charles~Lambert (\url{https://projets.lam.fr/projects/unsio/wiki}). } At the start of the simulation, the ICs are read and processed with help of the \dice\ \ramses\ patch \citep[][]{per16a}, modified to meet our specific needs, e.g. to read in additional particle arrays (e.g. {\sc part\_id}).

The methodology to calculate the cooling and heating of the gas has been described at length elsewhere \citep[e.g. sec.~2.2 in][]{Rey2020}. In brief, metallicity-dependent cooling is accounted for using the cooling functions by \citet{sut93a} for gas temperatures in the range $\log_{10} \left[{\rm T/ K} \right] = 4 - 8.5$, with rates from \citet{ros95b} used for cooling down to lower temperatures. Heating from a cosmic UV background is modelled following \citet{haa96a}, under the assumption that gas self-shields at high enough densities \citep[see][]{aub10a}.

The sub-grid prescriptions for star formation and stellar feedback have been developed in a series of papers \citep[][]{age09a,age09b,age13a,age15a,age15b,age16a,age21l}. In the following we offer a brief overview and refer the interested reader to the latter references for more details.

The galaxy-formation physics is an optional ingredient in our framework, which however becomes necessary if the aim is to create a synthetic galaxy with a realistic, multi-phase, turbulent interstellar medium (ISM; see below).

Star formation is treated as a Poisson process, sampled using $10^3~{\rm M}_\odot$ star particles, occurring on a cell-by-cell basis according to the star formation law $\dot{\rho}_{\star}= \epsilon_{\rm ff}\,\rho_{\rm g}/t_{\rm ff}$ for $\rho_{\rm g}>\rho_{\rm SF}$. Here $\dot{\rho}_{\star}$ is the star formation rate density, $\rho_{\rm g}$ the gas density, $t_{\rm ff}=\sqrt{3\pi/32G\rho_{\rm g}}$ is the free-fall time, and $\epsilon_{\rm ff}$ is the star formation efficiency per free-fall time of gas in the cell \citep[e.g.][]{fed12a}. The star formation threshold is set to $\rho_{\rm SF}=10(100)~{\rm cm}^{-3}$, depending on the application. We adopt a value of $\epsilon_{\rm ff}=10\,\%$ as this has been shown to give rise to realistic ISM and giant molecular cloud properties in Milky Way-mass disc simulations when coupled to our adopted feedback model \citep[][]{Grisdale2017,Grisdale2018}. 

Each formed star particle is treated as a single-age stellar population with a \citet{cha03a} initial mass function (IMF). Injection of energy, momentum, mass, and heavy elements over time from core-collapse supernovae (SNe) and supernovae Type Ia (SNIa), stellar winds, and radiation pressure into the surrounding gas is accounted for. Each of these mechanisms depends on stellar age, mass and gas/stellar metallicity, calibrated on the stellar evolution code {\small STARBURST99} \citep{lei99a}. The effect of supernova explosions are captured following the approach by \citet[]{kim15a}. Briefly, when the supernova cooling radius\footnote{The cooling radius in gas with density $n$ and metallicity $Z$ scales as \mbox{$r_\mathrm{cool} \approx 30 (n/1\cc)^{-0.43} (Z/Z_\odot + 0.01)^{-0.18} \pc$} for a SN explosion with energy $E_{\rm SN}=10^{51}$ erg.} is resolved by more than 6 grid cells, supernova explosions are initialised in the `energy conserving' phase by injecting $10^{51} \erg$ per SN into the nearest grid cell. When the cooling radius is resolved by less than grid 6 cells, the explosion is initialised in its `momentum conserving' phase, with the momentum built up during the von~Neumann-Sedov-Taylor phase \citep[][]{neu41a,sed46a,tay50a} injected into cells surrounding the star particle.\footnote{The adopted relation for the momentum is \mbox{$4\times~10^5 ~E_{\rm SN}/10^{51}\erg)^{16/17} (n/1~{\rm cm}^{-3})^{-2/17} (Z/Z_\odot)^{-0.2} \Msun\; \kms$} \citep[e.g.][]{kim15a,hop18a}, where $E_{\rm SN}$ is the total energy injected by supernovae (SNe) in a cell with gas density $n$ and metallicity $Z$ in solar units  ($\Zsun =0.02$).}

We track iron (Fe) and oxygen (O) abundances separately, with yields taken from \citet{woo07a}. When computing the gas cooling rate, which is a function of total metallicity, we construct a total metal mass \citep[cf.][]{kim14b}
$$M_{Z}=2.09M_{\rm O}+1.06M_{\rm Fe}$$
according to the mixture of alpha and iron group elements for the Sun \citep{asp09a}.\\

Future \nexus\ work (in particular at high numerical resolution) will benefit from a star-by-star treatment of star formation following the {\small INFERNO} model \citep[][]{Andersson2023}. Here, star particles masses are sampled from an IMF, hence representing \emph{individual} stars with their associated stellar evolutionary processes accounted for in a time-dependent manner. We will also consider a more extensive chemical treatment by tracking a wider range of elements (Andersson et al. in prep.), and we will compare predicted stellar abundance trends resulting from different choices of yield tables. It will also be advantageous to consider more sophisticated star formation models, including the impact of single- and multi-free fall models \citep[e.g.][]{fed12a} of the star formation rate per free fall time, adopted on a cell-by-cell basis.  

\section{Validation} \label{sec:val}

As a validation and a way of illustrating our framework, we set up a few test cases of general interest. We provide freely downloadable animations of the evolution of each of these test cases on our dedicated website: \url{http://www.physics.usyd.edu.au/nexus/movies/}. It is worth stressing that these cases are merely intended as control experiments, and not as realistic models of galaxies.

\paragraph*{A note on resolution} To limit the scope of the paper to the relevant aspects of the framework, we refrain for the time being from discussing what role the numerical resolution may play in the evolution of the idealised models that follow, and from performing the appropriate convergence tests. Instead, we defer this to future work, and adhere for the time being to criteria established by other groups about the appropriate resolution in terms of particle number required to avoid spurious effects such as artificial fragmentation of the gas \citep[][but see discussion in \citealt{van22a}]{tru97a}, discreteness effects \citep[][]{rom08a}, or numerical heating of collisionless components \citep[][]{wei98c,wil23a}.

\begin{table}
\begin{center}
\caption{ Model parameters common to all our models of a hot halo embedded in a responsive DM halo (see Sec.~\ref{sec:hothalo}). The total mass, scale length and cut-off radius are indicated in columns 2, 3, and 4 respectively. Column 5 gives the number of particles used to sample the corresponding component.
}
\label{tab:hhalos}
\begin{tabular}{llccccc}
Component & Mass & $r_s$ & $r_c$ & $N$ \\
 & ($10^{10}$ \Msun) & (kpc) & (kpc) & ($10^6$) \\
\hline
DM halo & 100 & 13.6 & 250 & 1 \\
Hot halo & 4.76 & 24.6 & 250 & 2 \\
\hline
\end{tabular}
\end{center}
\begin{list}{}{}
\item Notes. In the hydrostatic case (Sec.~\ref{sec:hadhalo}), the spin parameter of the hot halo is set to $\lambda = 0$; in all other cases, $\lambda = 0.12$; the DM halo does not have net rotation. Both components follow a \citet[][NFW]{nav97a} density profile (cf. Sec.~\ref{sec:gcorona_agama}). In the case of the cooling halo (Sec.~\ref{sec:schalo}), the initial metallicity is set to $5\times10^{-2}~\Zsun$.
\end{list}
\end{table}

\subsection{Hot halos} \label{sec:hothalo}

We start by looking at the simplest of models: a hot halo embedded within a responsive DM halo. The relevant parameters underlying all the following models are provided in Tab.~\ref{tab:hhalos}. Note that we have set the scale radius of the hot halo to be larger than that of the DM halo to emphasise the point that these components do not need to follow initially an identical structure.

\begin{figure}
\centering
\includegraphics[width=0.9\columnwidth]{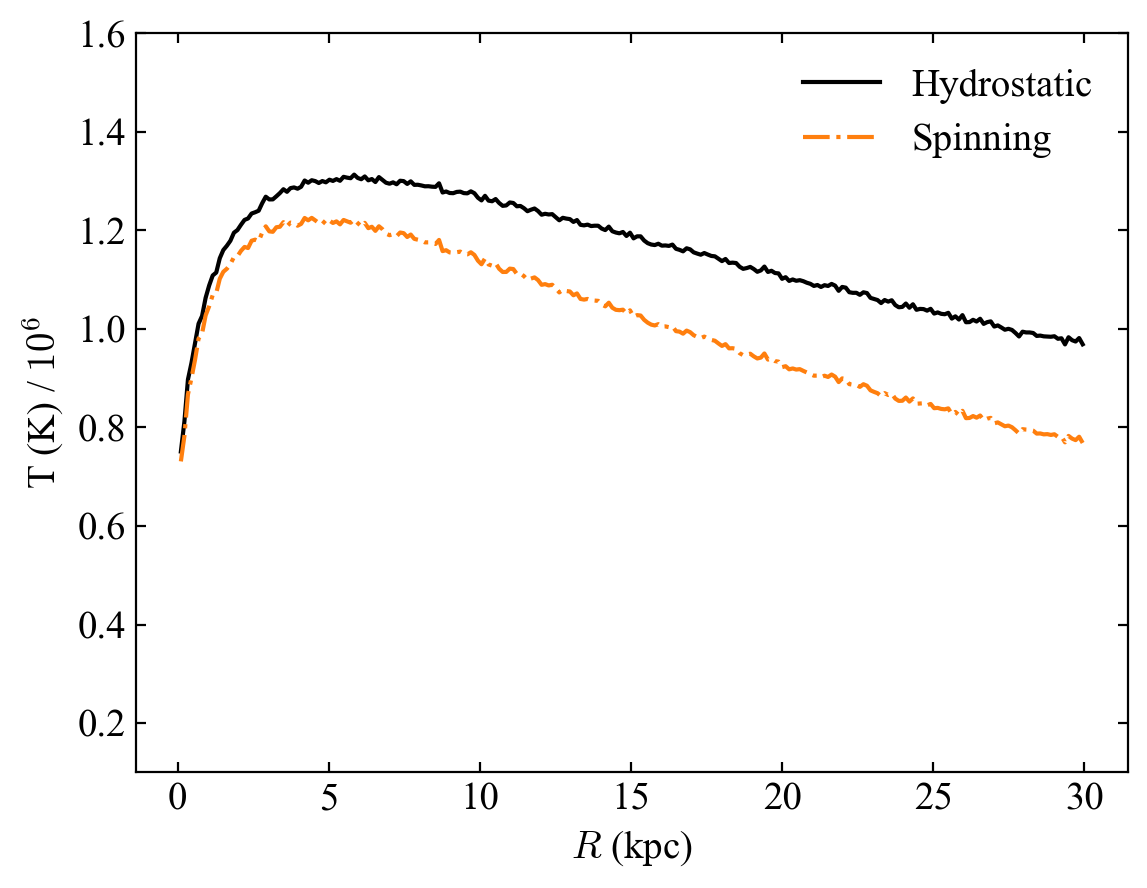}
\includegraphics[width=0.9\columnwidth]{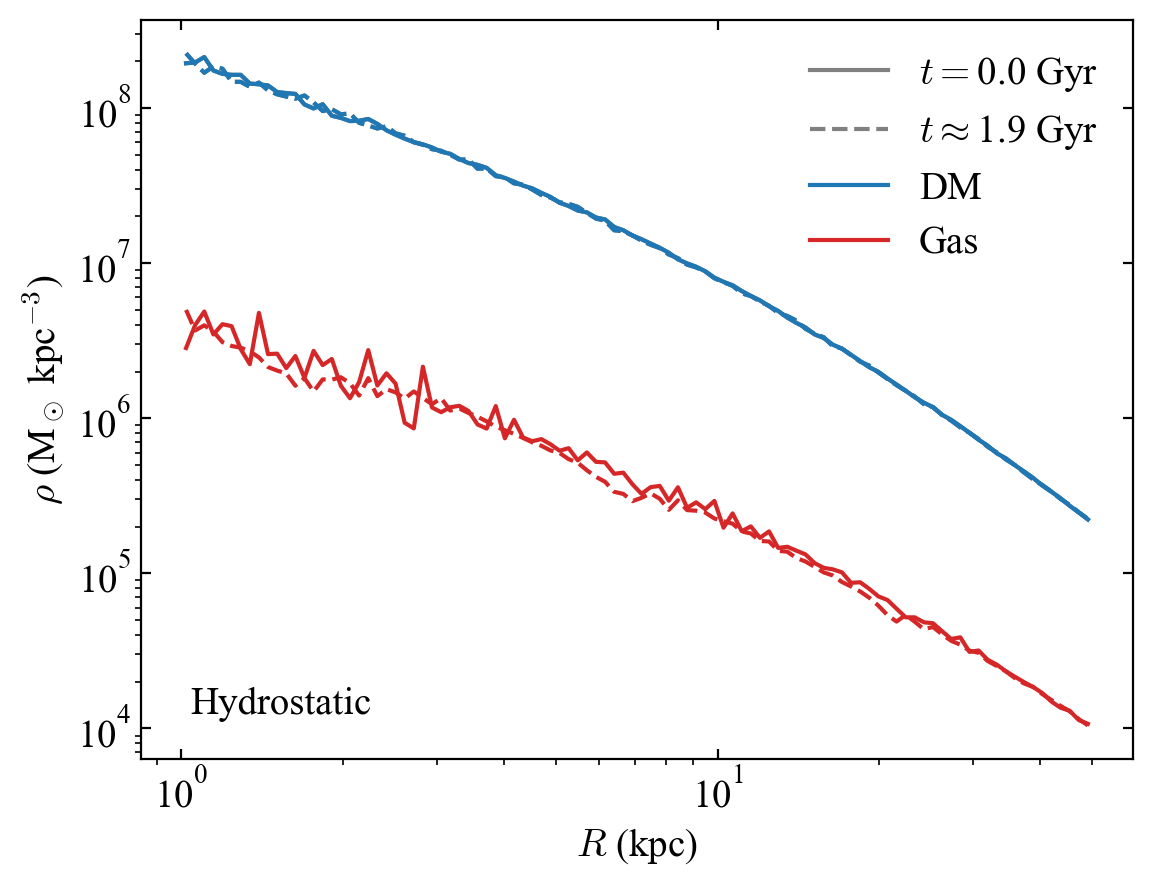}
\includegraphics[width=0.9\columnwidth]{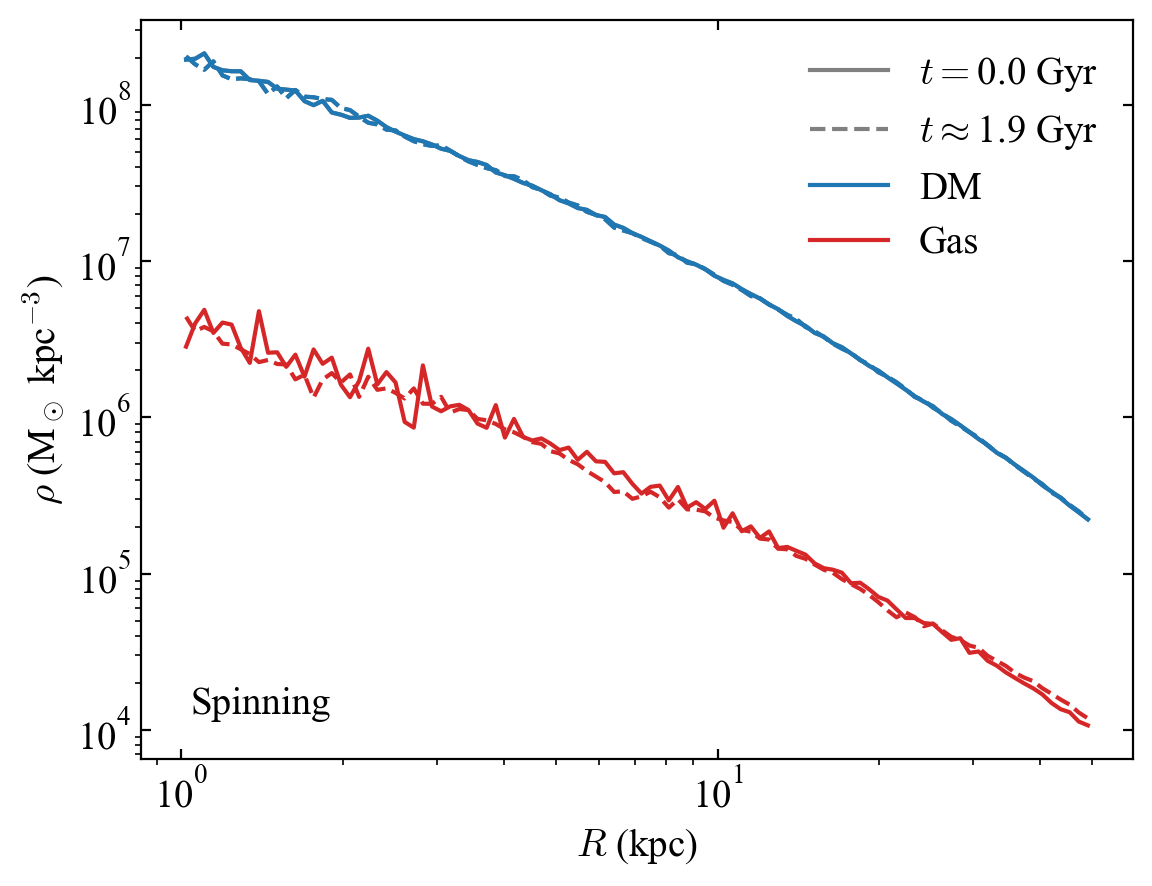}
\caption[  ]{ Top: Temperature profile of the hot halo in the hydrostatic case (solid black; Eq.~\ref{eq:thse}) and with spin (dot-dashed orange; Eq.~\ref{eq:trot}). In a state of equilibrium, the temperature of a spinning corona is lower relative to a pure hydrostatic configuration as a result of the additional rotational support against gravity. Middle/Bottom: Volume density profile of the DM halo (blue) and hot gas halo (red) in the hydrostatic case (middle) and the case of a spinning corona (bottom). The solid lines correspond in each case to the initial state of the component, and they are identical across panels. The dashed curve corresponds to the state of the component after roughly \mbox{$t~=~1.9$~Gyr} of evolution in isolation under adiabatic conditions. Note that the dashed lines, while not identical, are very similar across panels, and also very similar to their corresponding solid curves, attesting the stability of the initial conditions and the robustness of \ramses\ in evolving these.}
\label{fig:ghalo_props}
\end{figure}

\subsubsection{Hydrostatic, adiabatic halo} \label{sec:hadhalo}
Our first test case consists of a hydrostatic, hot halo in thermodynamic equilibrium with the total potential, with a temperature profile given by Eq.~\eqref{eq:thse}, and shown in the top panel of Fig.~\ref{fig:ghalo_props} by the solid, black curve. Clearly, the halo is not isothermal, but features a temperature on the order of $T \sim 10^6$~K -- close to the `virial temperature' of the DM halo -- with a profile that increases from the centre outwards up to \mbox{$r \approx 5$~kpc} only to decline again.

We evolve the composite system in a cubic box of size 500~kpc per side, adopting a maximum refinement level $l = 13$, implying a maximum spatial resolution of \mbox{500~kpc~/~$2^{13}~\approx~61$~pc}, for roughly 2~Gyr adopting an adiabatic equation of state.\footnote{An animation of the evolution of this model can be found at  \url{http://www.physics.usyd.edu.au/nexus/movies/h_00_gh1_lr_gas_xyz.mp4}. Note that there is virtually {\em no} change in the initial density distribution.} The expectation is that the system maintains its initial state indefinitely.

A departure from equilibrium is not trivial to quantify, but we can put an estimate based on e.g. the evolution of the density structure. We calculate the initial volume density profile of the hot halo and of the DM halo separately, and compare each to their corresponding profile at the end of the simulation. The result of this exercise is displayed in the central panel of Fig.~\ref{fig:ghalo_props}. Neither the DM halo (blue curve) nor the hot halo (red curve) display a significant evolution in terms of their mass distribution, as can be seen by comparing their corresponding initial profile (solid curves) with their profile at \mbox{$t~\approx~2$~Gyr} (dashed curves). In particular, neither component features a significant change in its central density \citep[generally indicative of the system being out of equilibrium; compare to e.g.][their figure 1]{tey13a}. The absence of such a change, and the general agreement of the initial and of the evolved profiles, are both strong indications that the system is in a stable, dynamical equilibrium from the outset.

The attentive reader may have noticed that the initial (\mbox{$t~=~0$}) density profile of the gas halo appears noisier in the central regions compared to the evolved (\mbox{$t~\approx~1.9$~Gyr}) density profile. The reason is that the initial density profile reflects the state of the ICs, which are affected by Poisson noise, as a result of sampling the profile with individual gas `particles' all with the same mass. For a monotonically declining density profile, the central regions are generally sampled with fewer particles compared to the outer regions, and are thus noisier. At initialisation, \ramses\ maps the gas particles onto the AMR grid, which is refined at runtime (\mbox{$t~>~0$}) according to a number of criteria. One of these is to refine cells with density above a prescribed threshold. Since the central regions are denser compared to the outer regions, they are refined up to higher levels, leading to a smoother density distribution compared to the initial one, provided the gas profile {\em does not evolve}.\\

We have made use of this type of halo model extensively in the past \citep[][]{tep18a,tep18b,tep19a}. Despite such models being somewhat simplistic, they have proven useful as demonstrated by the latter study in particular, which emphasised the importance of the presence of a hot halo component in galaxy models that push towards completeness, and which ultimately led to the prediction \citep{luc20a} and putative detection \citep{kri22a} of the Magellanic Corona.

\begin{figure*}
\centering
\includegraphics[width=\textwidth]{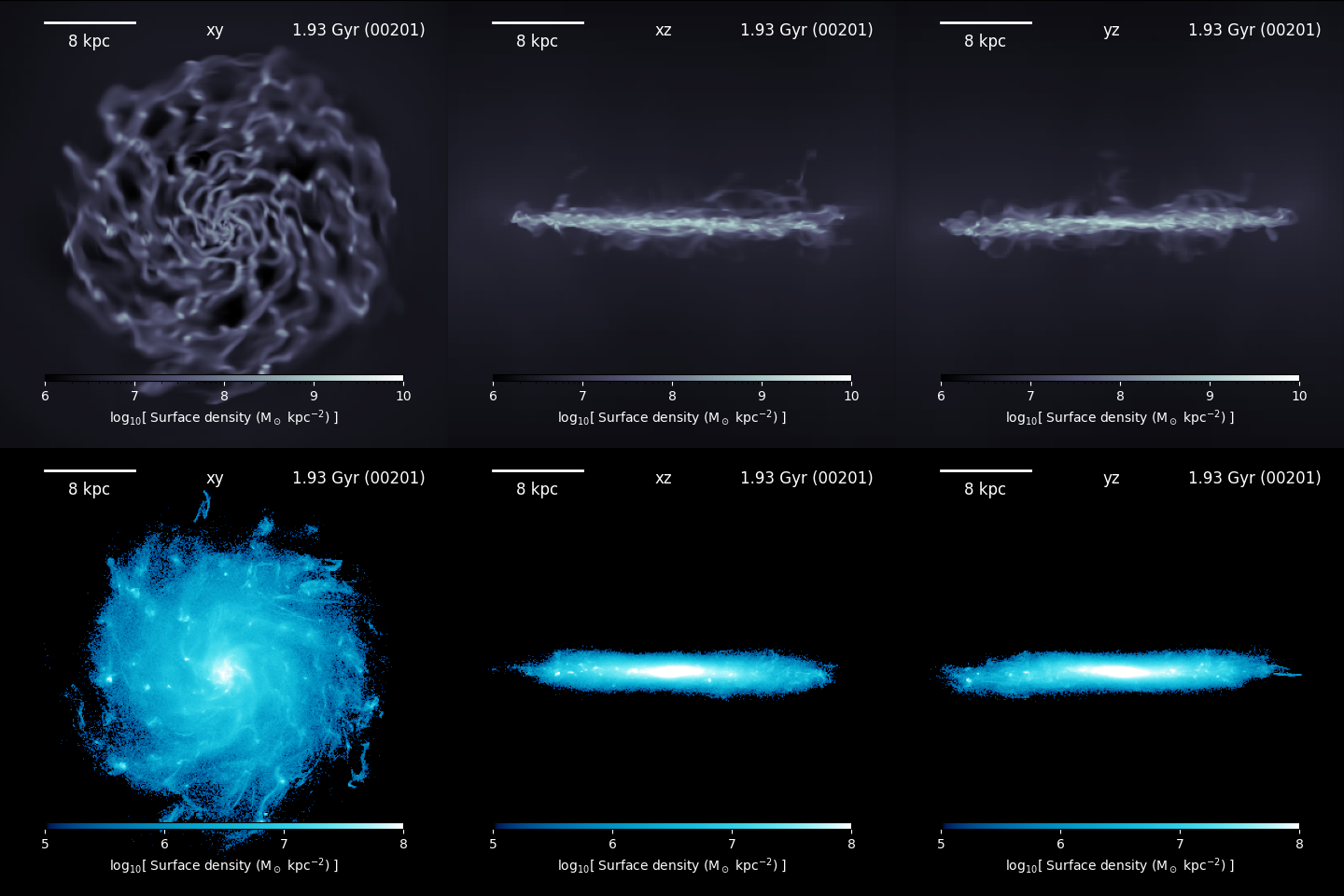}
\caption[  ]{ State of a spinning, cooling halo embedded within a responsive DM halo after roughly 2~Gyr of evolution. Top: Gas. Bottom: Stars }
\label{fig:ghalo_evol}
\end{figure*}

\subsubsection{Spinning, adiabatic halo} \label{sec:sadhalo}

Our next test setup consists of a non-isothermal, spinning corona, in vertical hydrostatic equilibrium. Its temperature profile is shown in the top panel of Fig.~\ref{fig:ghalo_props} by the dot-dashed, orange curve. The mass distribution of this model is identical to the non-rotating, hydrostatic model discussed in Sec.~\ref{sec:hadhalo}, but their temperature profiles are clearly different as a consequence of the different kinematic structure. Specifically, this spinning model features an overall lower temperature profile, because the rotation velocity provides additional support against gravity (Eq.~\ref{eq:trot}). The velocity has been set by requiring that the spin parameter of the hot halo be \mbox{$\lambda~=~0.12$}, consistent with the findings from cosmological simulations (Sec.~\ref{sec:ghvel}). Note that the DM halo has no net rotation in any of our models (but it is worth emphasising that it can be easily added if so desired within our framework).

As with the non-rotating, hydrostatic model, the spinning corona model is expected to be in a stable equilibrium from the outset. To check for this, we proceed as we did with the non-rotating, hydrostatic model: we calculate the evolution of the systems over roughly 2~Gyr under adiabatic conditions, and compare the initial and final density profiles of each component (DM halo, hot halo). The result is displayed in the bottom panel of Fig.~\ref{fig:ghalo_props}. Overall, the final density profiles of both components agree well with their respective initial profile. Based on this, we are confident that the system is in a stable equilibrium from the outset.\footnote{An animation of the evolution of this model can be found at \url{http://www.physics.usyd.edu.au/nexus/movies/h_00_gh0_lr_ad_gas_xyz.mp4}.}

\subsubsection{Spinning, cooling halo} \label{sec:schalo}

The last test within the category of hot halos consists in the classical setup of a spinning, cooling galactic halo (or `cooling halo' for short) pioneered by \citet[][see also \citealt{Noguchi1999}]{kau06a}, followed by a long list of studies of both isolated systems \citep[e.g.][]{ros08v,tey13a,hob13a,mar15a,kho21a}, and merging systems \citep[e.g.][]{hwa15a,ath16a}.

The initial conditions are identical to the model discussed in Sec.~\ref{sec:sadhalo}, but they are advanced in time, allowing the gas to evolve thermodynamically (cool / heat) and to form stars (Sec.~\ref{sec:galevol}). In this case, the system will not retain its initial configuration. Rather, the hot gas loses part of its pressure support as a result of cooling, and collapses. Angular momentum conservation deters the gas from collapsing spherically, and collapse proceeds along the spin axis, resulting in a disc-like, rotationally supported gas configuration.

The state of the system after roughly 2~Gyr of evolution is presented in Fig.~\ref{fig:ghalo_evol}. The top row displays the gas distribution along three orthogonal projections: face-on (left), and side-on (centre~/~right). The bottom row shows the distribution of stars along the same projections. The gas disc features a series of flocculent spiral arms, as well as gas plumes and other gas structures reminiscent of galactic fountains. The stelllar disc appears thickened; it displays spiral arm-like features, and a number of dense stellar `knots', including a central mass concentration. These results are very much in agreement with the results of similar earlier work (see references above).\footnote{An animation of the evolution of this model can be found at \url{http://www.physics.usyd.edu.au/nexus/movies/h_00_gh0_lr_ofe_wh_xyz.mp4}.}

This classical setup is indeed a beautiful demonstration of the idea that galaxies may form out of the cooling of shock-heated, intergalactic gas accreted onto DM halos (cf. Sec.~\ref{sec:gcorona}).

\begin{figure*}
\centering
\includegraphics[width=\textwidth]{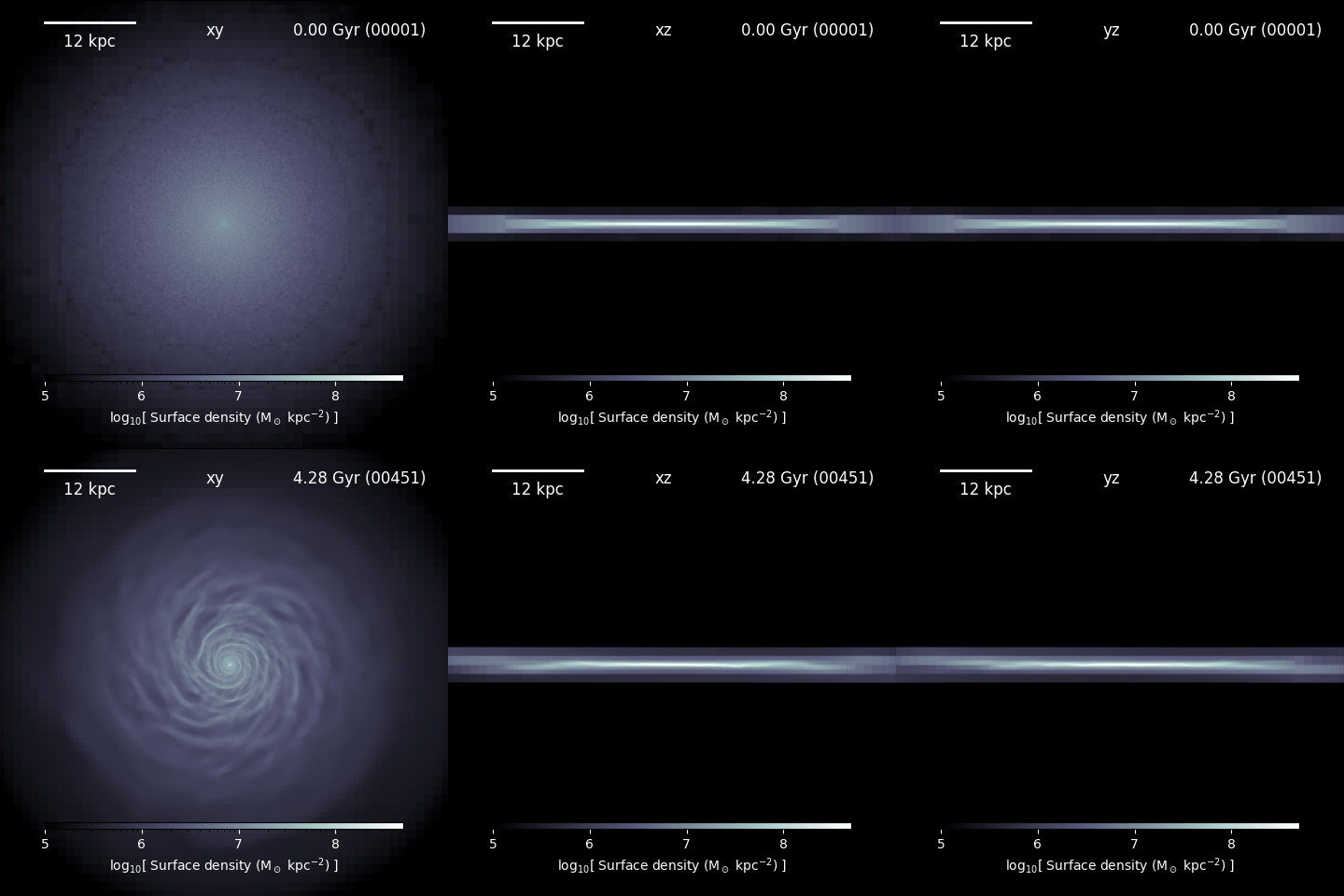}
\caption[  ]{Gas surface density of a strict isothermal disc embedded within a responsive DM halo -- bulge -- stellar disc system (not shown). Top: Initial state. Bottom: State after roughly 4.3~Gyr of evolution. Note that the disc develops sub-structure, clearly visible on the face-on projection (left), as well as a pile-up of mass at the centre (see also Fig.~\ref{fig:gdisc_sdens}), but it retains its overall structure, in particular its thickness (middle, right).}
\label{fig:gdisc_evol}
\end{figure*}

\begin{figure}
\centering
\includegraphics[width=\columnwidth]{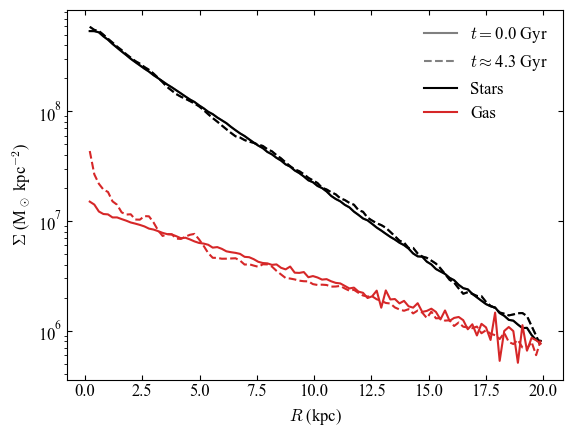}
\caption[  ]{ Surface density profile of the stellar disc (black) and gas disc (red) of the simulation test shown in Fig.~\ref{fig:gdisc_evol}. The solid lines correspond in each case to the initial state of the component. The dashed curve corresponds to the state of the component after roughly 4.3~Gyr of evolution in isolation using an isothermal EoS. Note that the dashed lines, while not identical, are very similar to their corresponding solid curves, attesting the stability of the initial conditions and the robustness of \ramses\ in evolving these. The most notable difference between these two states is the pile-up of gas mass close to the centre (see also Fig.~\ref{fig:gdisc_evol}).
}
\label{fig:gdisc_sdens}
\end{figure}

\begin{table*}
\begin{center}
\caption{Galaxy model parameters (Sec.~\ref{sec:tdisc}). Columns 1 and 2 identify the galactic components and their associated functional forms; we note that these are approximations because they share the same gravitational potential. The total mass, scale length and cut-off radius are indicated in columns 3, 4, and 5 respectively. Column 6 is the number of particles used to sample the corresponding component. Note that for gaseous components, this number corresponds only to the initial particle number.
}
\label{tab:hbd_10_gd2}
\begin{tabular}{llccccc}
Component & Profile & Total mass & Radial scalelength & Cut-off radius & Particle count \\
 &  & $M_{\rm tot}$  & $r_s$ & $r_c$ & $N$ \\
 &  & ($10^{10}$ \Msun) & (kpc) & (kpc) & ($10^5$) \\
\hline
DM halo & NFW & 145 & 15 & 300 & 20 \\
Stellar bulge & Hernquist & 1.5 & 0.6 & 2.0 & 4.5 \\
Stellar disc & Exp, $\sech^2$ & 3.4 & 3.0 & 40 & 10\\
Gas disc & Exp, $\sech^2$ & 0.4 & 7.0 & -- & 20\\
\hline
\end{tabular}
\end{center}
\begin{list}{}{}
\item Notes: The NFW and Hernquist functions are defined elsewhere \citep[][]{nav97a,her90a}. The scaleheight of the stellar disc is $z_d \approx 250$ pc; the Toomre local instability parameter of the stellar disc \citep{too64a} is everywhere $Q \gtrsim 1.3$. The gas disc is isothermal with \mbox{$T~=~10^3$~K}, with a scaleheight that varies with radius from roughly 20 pc near the centre to 160 pc at \mbox{$R~=~20$~kpc} (a `flaring' disc). The gas disc is not truncated, but merges smoothly with the background density (set at \mbox{$10^{-20}$~\pcc} in our \ramses\ setup).
\end{list}
\end{table*}

\subsection{A galaxy with an isothermal gas disc} \label{sec:tdisc}

As a final test of our framework, we set up an isothermal (\mbox{$T = 10^3$~K}) gas disc embedded in a responsive DM Halo-Bulge-disc (HBD) system, i.e. an isolated galaxy consisting of a DM halo, a classical stellar bulge, a stellar disc, and a gaseous disc. This model is virtually identical to the isolated galaxy model underlying one of our earlier studies \citep{tep22x}, but at a somewhat lower particle resolution. The relevant model parameters are displayed in Tab.~\ref{tab:hbd_10_gd2}.

We evolve the composite system in a cubic box of size 600~kpc per side, adopting a maximum refinement level $l~=~14$, implying a maximum spatial resolution of \mbox{600~kpc~/~$2^{14}~\approx~37$~pc}, for roughly 4.3~Gyr using a strict isothermal equation of state.\footnote{An animation of the evolution of the stellar disc (blue) and gaseous disc (orange) in this model can be found at \url{http://www.physics.usyd.edu.au/nexus/movies/hbd_10_gd2_xyz.mp4}.\label{foo:gdisc}}

Under these conditions, we expect the system to retain its initial state. To test this expectation. we compare the initial surface density profile of the stellar disc and of the gas disc to their corresponding profile after 4.3~Gyr of evolution. We do not look at the DM halo and bulge, as we have shown previously that such spheroidal components remain roughly in a stable equilibrium within our framework.

A snapshot of the initial state gas disc and its state at \mbox{$t~\approx~4.3$~Gyr} is displayed in the top and bottom panels of Fig.~\ref{fig:gdisc_evol}, respectively. Although the disc clearly departs from its initially smooth appearance and develops sub-structure, it retains its overall shape, in particular its thickness. This is a clear improvement over the results from similar approaches \cite[compare to e.g.][their figs.~1 and 6]{deg19a}. The stellar disc displays a similar behaviour (not shown, but see Footnote \ref{foo:gdisc}).

A more quantitative assessment of the system's stability is presented in Fig.~\ref{fig:gdisc_sdens}, which displays the initial surface density profile of the stellar disc (solid black curve) and of the gas disc (solid red curve), and their corresponding profile after roughly 4.3~Gyr of evolution (black and red dashed curves, respectively). The stellar disc shows barely any change with respect to its initial state. The gas disc does show some departure from its initial state, notably a mass pile-up close to the centre and instabilities at the edge. Nonetheless, overall, the system appears to be in a stable dynamical equilibrium from the outset.

\section{A nested-bar system} \label{sec:s2b}
 
As a further application of our framework, we extend the model of a barred MW surrogate previously discussed in \citet{tep21v} to include a gaseous disc component. The relevant model parameters are listed in Tab.~\ref{tab:s3b}. The synthetic galaxy thus consists of four components, all of them responsive: a DM host halo, a pre-existent stellar bulge, a pre-existent stellar disc, and a gas disc. The disc-to-total mass fraction of the model is \mbox{$f_\mathrm{disc}~\approx~0.45$}, which renders the disc bar-unstable from the outset \citep[cf.][]{fuj18a,bla23a}.

\begin{table*}
\begin{center}
\caption{Galaxy model parameters (cf. Sec.~\ref{sec:s2b}). Meaning of columns and quantities identical to Tab.~\ref{tab:hbd_10_gd2}.
}
\label{tab:s3b}
\begin{tabular}{llccccc}
Component & Profile & Total mass & Radial scalelength & Cut-off radius & Particle count \\
 &  & ($10^{10}$ \Msun) & (kpc) & (kpc) & ($10^6$) \\
\hline
DM halo & NFW & 118 & 19 & 250 & 1 \\
Stellar bulge & Hernquist & 1.25 & 0.6 & 2.0 & 0.1 \\
Stellar disc & Exp, $\sech^2$ & 4.31 & 2.5 & 25 & 1 \\
Gas disc & Exp, $\sech^2$ & 0.46 & 3.5 & -- & 2 \\
\hline
\end{tabular}
\end{center}
\begin{list}{}{}
\item Notes: The scaleheight of the stellar disc is \mbox{$z_d~\approx~300$~pc}; the Toomre local instability parameter of the stellar disc \citep{too64a} is everywhere $Q \gtrsim 1.3$. The gas disc is initially isothermal with $T = 10^3$ K. The initial gas metallicity is set to \mbox{$Z~=~1$~\Zsun} in the disc, and to zero elsewhere. The gas disc merges smoothly with the background density (\mbox{$\rho~=~10^{-20}$~\pcc}, \mbox{$T~=~10^6$~K}). The disc-to-total mass fraction is \mbox{$f_{\rm disc}~\approx~0.45$}, which renders the disc bar-unstable from the outset \citep[cf.][]{fuj18a,bla23a}.
\end{list}
\end{table*}

\subsection{Simulation and results}

We evolve the initial conditions for roughly 4~Gyr in a cubic volume of 600~kpc across, adopting a maximum refinement level $l = 13$, impying a limiting spatial resolution of roughly 73~pc. The volume is filled with an additional hot (\mbox{$T~=~10^6$~K}), tenuous (\mbox{$n~\leq~10^{-6}~\pcc$}) gas atmosphere. The self-gravity of all components is taken into account, and the gas is allowed to evolve thermodynamically and to form stars (cf. Sec.~\ref{sec:galevol}).

Given the relatively high central gas densities, star formation takes place there nearly instantly and vigorously, leading to the formation of a young stellar disc that grows in an inside-out fashion. At the same time, the pre-existent stellar disc succumbs to the bar instability and develops a clear bar-like structure at the centre at \mbox{$t~\lesssim~1$~Gyr}.

The young stellar disc follows suit, and develops a bar of its own aligned with the pre-existent stellar bar of roughly the same size. In addition, the young stars spectacularly develop a {\em second, inner} bar. Thus, we distinguish between an outer, pre-existent stellar bar, an outer, newly formed stellar bar, and an inner, newly formed stellar bar.\footnote{An animation of the evolution of this model showing the evolution of the pre-existent stellar disc (blue-white), of the gaseous disc (orange), and the newly formed stellar disc (blue-yellow) can be found at \url{http://www.physics.usyd.edu.au/nexus/movies/hbd_11_gd9_lr_star_xyz.mp4}.} 

\begin{figure*}
\centering
\includegraphics[width=0.33\textwidth]{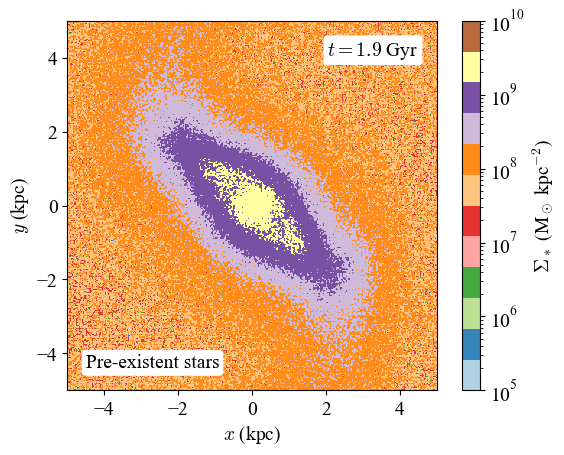}
\includegraphics[width=0.33\textwidth]{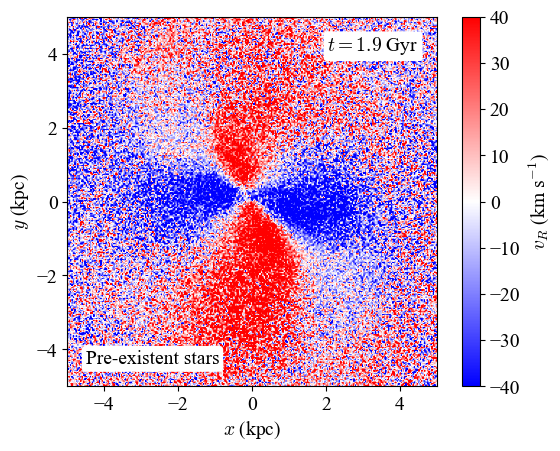}
\includegraphics[width=0.33\textwidth]{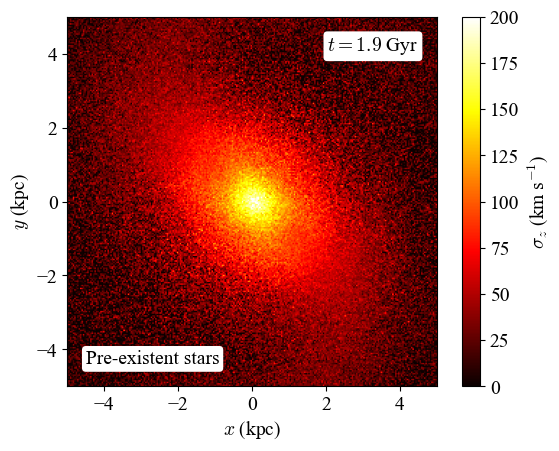}
\includegraphics[width=0.33\textwidth]{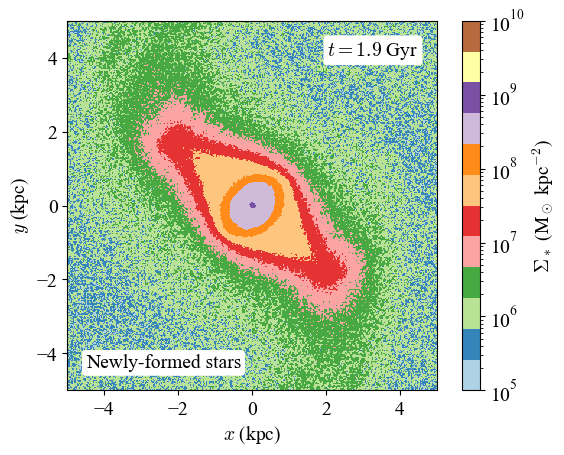}
\includegraphics[width=0.33\textwidth]{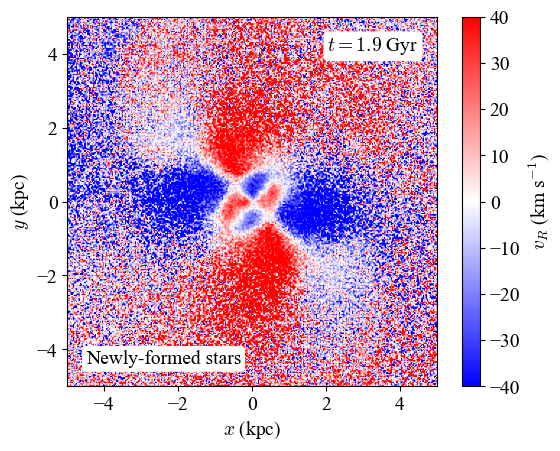}
\includegraphics[width=0.33\textwidth]{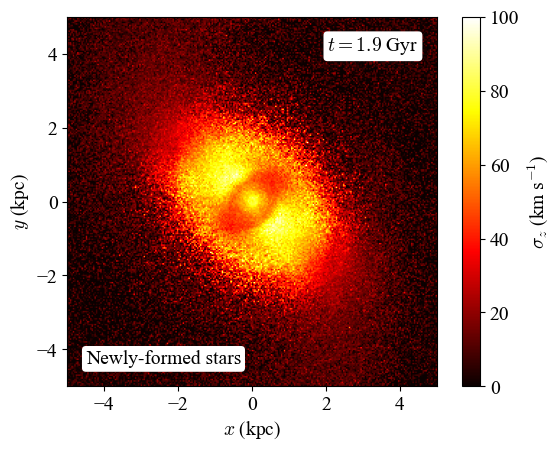}
\includegraphics[width=0.33\textwidth]{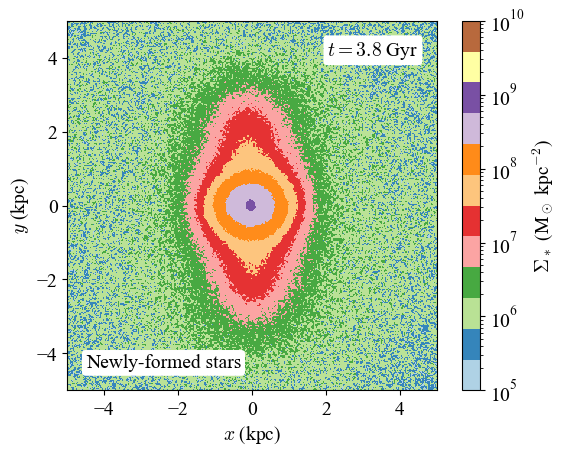}
\includegraphics[width=0.33\textwidth]{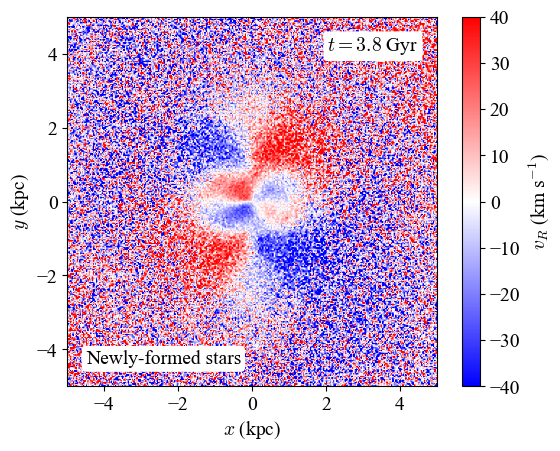}
\includegraphics[width=0.33\textwidth]{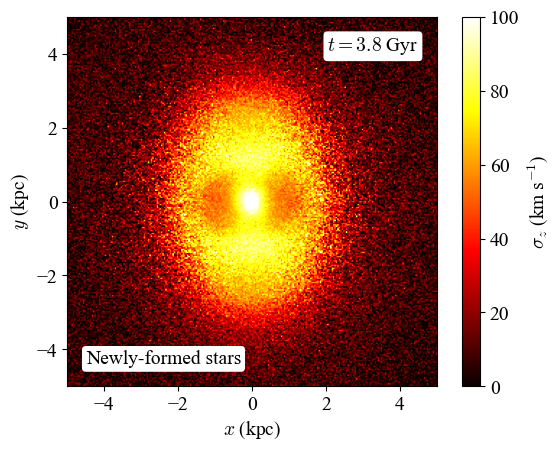}\\[5pt]
\caption[  ]{ Projected mass density (left), projected radial velocity (centre), and projected vertical velocity dispersion (right) of the pre-existent stars (top), and of the newly formed stars (middle/bottom) after \mbox{$t~=~1.9$~Gyr} (top/middle) and \mbox{$t~=~3.8$~Gyr} (bottom) of evolution. Note the nested-bar structure in the young stellar component. The outer bar aligns with the bar in the pre-existent stellar component, and precedes the formation of the inner bar (cf.~Fig.~\ref{fig:nested_bars_age_met_sfr}), which appears perpendicular to the outer bar. Notice the difference in scale in the vertical velocity dispersion between the top and the bottom rows, so chosen to avoid colour saturation. The galaxy shows the characteristic kinematic features of nested-bar galaxies: the double quadrupole for mean $v_R$ (centre) and humps at the minor axis of the inner bar for $\sigma_z$ (right). The central part of the inner bar appears to have collapsed into a bulge by the end of the simulation. Note that we have not corrected for the bar phase across snapshots.}
\label{fig:nested_bars}
\end{figure*}

In Fig.~\ref{fig:nested_bars}, we show the state of the pre-existent stellar disc (top row) and of the newly formed stellar disc (middle row) after \mbox{$t~=~1.9$~Gyr} of evolution within a region enclosed by \mbox{$(x,y)~\in~[-5,5]\times[-5,5]$~kpc}. In each case, the left column displays the projected stellar density; the central column displays the mean projected radial velocity of the stars ($v_R$), and the last column displays the projected vertical velocity dispersion of the stars ($\sigma_z$).

Focusing on the first row, it is clear that the pre-existent stellar disc has lead to the formation of a central bar with a declining density profile (top-left column), and the classical quadrupole pattern $v_R$ (top-middle column), and a vertical velocity dispersion $\sigma_z$ that declines from the centre outwards (top-right column). At the same epoch, the newly formed stellar disc (middle row) also displays a bar-like structure at its centre, which appears aligned with the bar in the pre-existent stellar disc, but is somewhat weaker in terms of density (middle-left column). A notable difference between the pre-existent and the newly formed bars is the presence in the latter of an oblong, central mass concentration with a semi-mayor axis seemingly perpendicular to that of the outer bar. The presence of this inner structure is apparent in the $v_R$-map (middle-central column), which displays two quadrupole patterns, one smaller embedded within the more extended one. This is reminiscent of the kinematics observed in a similar system found in cosmological simulations, albeit formed by tidal interaction rather than by an internal instability \citep[][their fig.~4]{sem24a}.

The vertical velocity dispersion of the newly formed disc (middle-right column) differs significantly from that of the pre-existent stellar disc. Notably, it displays two kinematically distinct components: one with a kinematic structure reminiscent of the pre-existent stellar bar, but less extended, which appears to align with the inner bar, superimposed on a more extended structure with a hot kinematic signature along its semi-major axis, aligned with the outer bar. These features are the so-called $\sigma$-`humps' and $\sigma$-`hollows' previously identified by \citet[][see also \citealt{de-08a}]{du16a}, and reproduced in some simulations \citep[e.g.][]{li23b,sem24a}.

The last row in Fig.~\ref{fig:nested_bars} displays the state of the system after \mbox{$t~=~3.8$~Gyr}. A few differences with respect to the state of the newly formed disc roughly 2~Gyr earlier (middle row) are apparent. First, the central part of the inner bar appears to have collapsed into a bulge by the end of the simulation (bottom-left column). This is interesting because it has been suggested that dissolved inner bars may be the origin of classical bulges \citep[][]{du17a}. Moreover, the $v_R$ map is less well-defined, in particular within the inner bar region, suggesting that the bars may be weakening. Finally, while the $\sigma_z$ map displays the same qualitative features (hollows, humps) it suggests that the bars have become kinematically hotter.

\begin{figure*}
\centering
\includegraphics[width=0.48\textwidth]{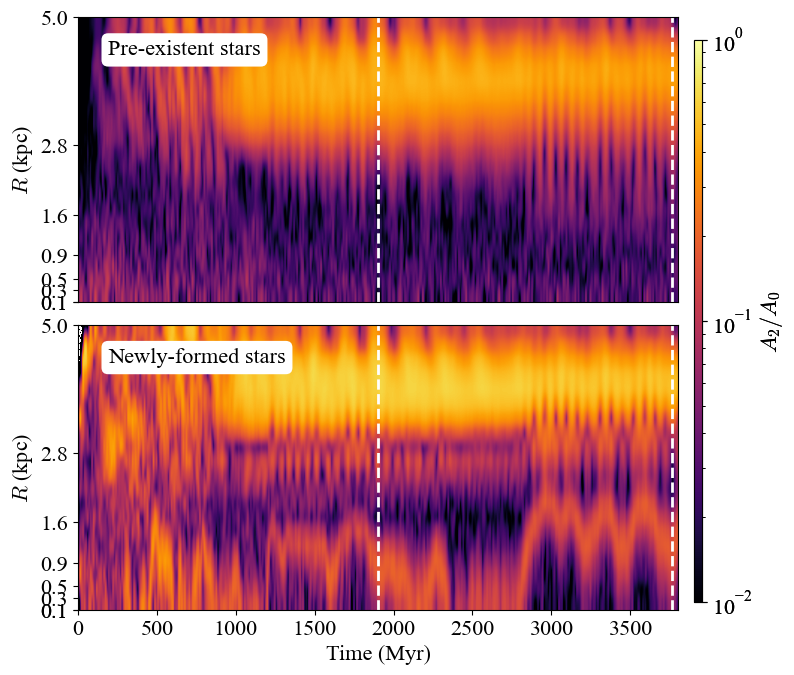}
\includegraphics[width=0.48\textwidth]{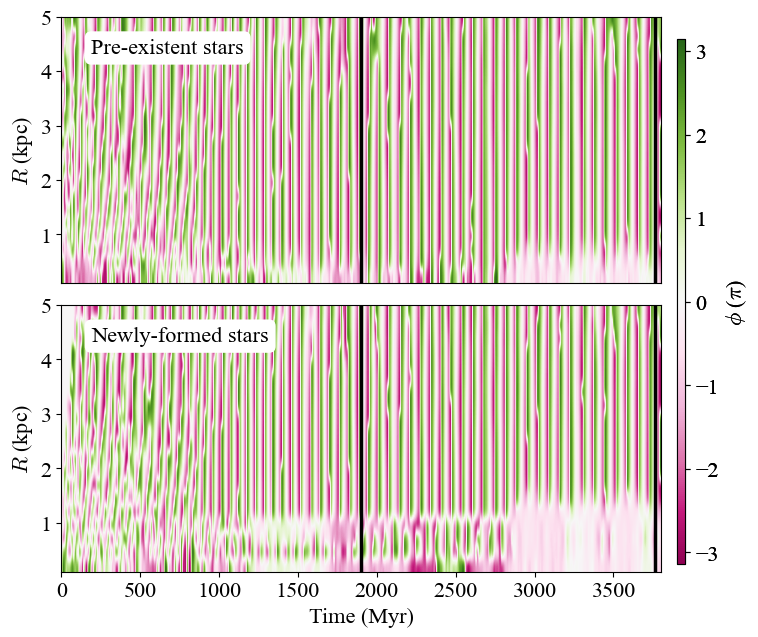}
\caption[  ]{ Evolution of the $m~=~2$ Fourier amplitude ($A_2~/~A_0$; left) and its phase ($\phi$; right) across the bar region ($R~\leq~5$~kpc) over a period of roughly 3.8~Gyr for each stellar component separately. Top: Pre-existent stars. Bottom: Newly formed stars. Note that the radial range is equally spaced on a logarithmic scale on the left panels, but on a linear scale in the right panels. The bottom panel displays a clear strong signal (i.e. a high $A_2~/~A_0$ value as indicated by the colour-bar on the right) both within \mbox{$0~<~R/\kpc~\lesssim~2$} and \mbox{$3~\lesssim~R/\kpc~<~5$}, indicating the presence of an inner and of an outer bar, respectively. The top panel indicates the presence of an outer bar only. The inner bar phase appears in lock-step with the outer bar, suggesting they remain perpendicular at all times. Note that all bars persist out to at least \mbox{$t~\sim~2.7$~Gyr}, after which it seems to lead to the formation of a central mass concentration (likely a classical bulge). The vertical lines (left: dashed-white; right: solid-black) flag the epoch corresponding to the snapshots underlying Figs.~\ref{fig:nested_bars} and \ref{fig:nested_bars_age_met_sfr}.  }
\label{fig:a2-phase}
\end{figure*}

We look at the behaviour and structure of the stellar discs throughout the simulation in a holistic way by calculating the evolution of the amplitude of the $m~=~2$ Fourier mode, \mbox{$\overline{A}_2~\equiv~A_2~/~A_0$} and its phase $\phi[\overline{A}_2]$ in space and time \citep[e.g.][]{mit23a}. In practice, we calculate the radial profile of $\overline{A}_2$ in radial bins \mbox{$\Delta R = 0.5$~kpc} within a circular aperture with radius \mbox{$R~=~5$~kpc} at every time step \mbox{$\Delta t~=~10$~Myr}, effectively yielding a map $\overline{A}_2(R,~T)$ and $\phi(R,T)$, respectively, for each of the pre-existent stellar disc and the newly formed stellar disc. The result of this exercise is shown in Fig.~\ref{fig:a2-phase}.

The left panels display $\overline{A}_2(R,~T)$ for the pre-existent stellar disc (top) and the newly formed stellar disc (bottom). The formation of a bi-symmetric structure within \mbox{$R~=~5$~kpc}, signalled by a high value of $\overline{A}_2$, is apparent in the pre-existent stellar disc at \mbox{$t~\approx~1$~Gyr}, persisting all the way to \mbox{$t~\approx~3.9$~Gyr}. The same is true for the newly formed stellar disc. It, however, also displays an inner ($R~\lesssim~2$~kpc) signal with a high  $\overline{A}_2$ amplitude indicative of the presence of an inner bar. Its signal is less consistent than that of the outer bar, and the signal close to the centre weakens at \mbox{$t~\gtrsim~2.7$~Gyr}, suggesting the bar is slowly dissolving. This is consistent with the density distribution appearing more centrally concentrated towards the end of the simulation (cf. Fig.~\ref{fig:nested_bars}, bottom-left). 

The right panels display $\phi(R,T)$ for each of the pre-existent stellar disc (top) and the newly formed stellar disc (bottom). The former shows an alternating pattern in $\phi$, as expected for a rotating bar, which appears constant all the way to the centre at a given time step, strongly supporting the existence of a well-defined pattern speed of the outer pre-existent bar. This behaviour is mimicked by the newly formed stellar disc. In contrast, there is a second signal, most apparent within \mbox{$R~=~1$~kpc}, which presumably indicates the pattern speed of the inner bar. It seems to be in lock-step with the outer bar, suggesting their semi-major axes are and remain perpendicular to one another at all times. Note that the signal is fading at \mbox{$t~\gtrsim~2.7$~Gyr}, consistent with the weakening of $\overline{A}_2$ at the same epoch.

We note the `pulsating' nature of the $\overline{A}_2$ amplitude in both disc components, which shows an abrupt change at \mbox{$t~\approx~2.8$~Gyr} in both components. Pure visual inspection suggests that its `beat' is different from the phase's beat, and it is unclear at this point whether they are related and what the origin of the former might be.

~\\
Given the putative connection between the presence of nested bars in galaxies and the fuelling of active galactic nuclei (AGN) and black-hole (BH) growth \citep{shl89a,nam09a,du17a}, as well as vigorous star formation activity \citep[][]{rom16a}, it is interesting to look now at the gas flow in and around the galaxy centre. To this end, we proceed as follows. First, we calculate the enclosed mass within two cilindrical apertures with height \mbox{$| z | \leq 1$~kpc} and radii \mbox{$R~=~5$~kpc} and \mbox{$R~=~2$~kpc}, roughly enclosing the outer bars and the inner bar. We separately look at the enclosed gas mass and the mass of newly formed stars, as well as at their combined mass (since stars form from gas). The latter approach allows assessing the mass change within the respective bar regions. To gain some insight into the mass flow, we calculate the time derivative of the enclosed mass profile, following \citet{li23b}.

\begin{figure*}
\centering
\includegraphics[width=0.49\textwidth]{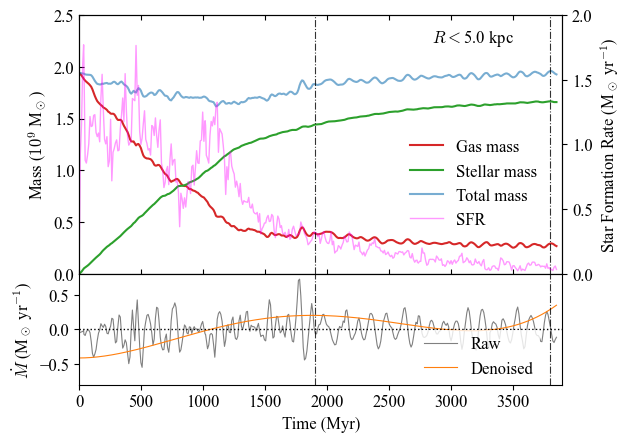}
\includegraphics[width=0.49\textwidth]{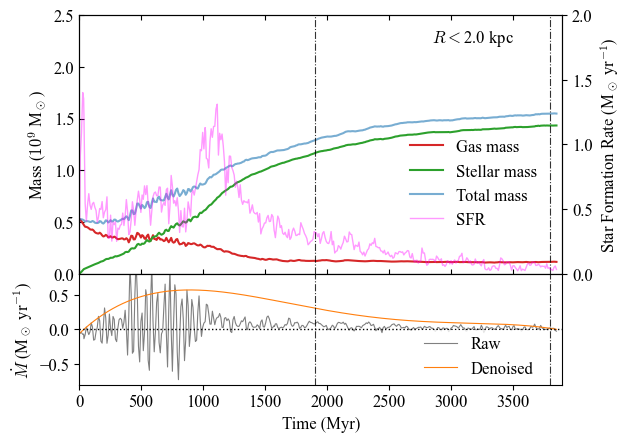}
\caption[  ]{ Evolution of the mass (gaseous, stellar, and total, i.e. gaseous and stellar added together) enclosed within two circular apertures: \mbox{$R~<~5$~kpc} (left) and \mbox{$R~<~2$~kpc} (right). The total star-formation rate within each of these regions is shown as well (scale is provided on the right $y$-axis and is identical in both panels). The sub-panel at the bottom displays in each case the total mass flow rate, estimated from the derivative with respect to time of the total enclosed mass (see text for details). The vertical, dot-dashed lines flag the epoch corresponding to the snapshots underlying Figs.~\ref{fig:nested_bars} and \ref{fig:nested_bars_age_met_sfr}. }
\label{fig:mass}
\end{figure*}

The result is presented in Fig.~\ref{fig:mass}. The main panels display the gas (red curve), stellar (green curve) and total (gas plus stellar; blue curve) mass enclosed within two circular apertures around the centre with radius \mbox{$R~=~2$~kpc} (left) and 5~kpc (right). The sub-panels in each case display the total mass flow rate, $\dot{M}$ (in units of \Msun~\pyr) estimated from the derivative with respect to time of the total enclosed mass. The derivative is calculated both on the total enclosed mass data as is (termed `raw'; grey curve) and a denoised version of it (orange curve). The latter is calculated by interpolating the raw data with a spline function.\footnote{We accomplish this with the help of the {\tt splrep} and the {\tt splev} modules (setting $k~=~5$ and $s~=~3$), both provided by the {\small SCIPY} package \citep{vir20a}.} The reason is that the derivative of the raw data may provide information on the periodicity of the mass flow, while the derivative of the denoised data better estimates the actual net mass flow.

We find that the inner region, which encloses the inner bar, experiences a net mass inflow, while the outer region -- which encloses the outer bar -- displays a rather flat mass growth on average. This behaviour can be understood by looking at the star-formation rate (SFR) averaged over the aperture, shown by the magenta curve, in each case. Within the larger region (right panel), we observe a clear correlation between enhanced SFR, gas depletion (red), and stellar mass growth (blue) over a period of \mbox{$t~\approx~1.5$~Gyr} since the start of the simulation. Overall, the gas depletion dominates over stellar mass growth ($\dot{M}~<~0$, orange curve) during that time, suggesting that some of the gas is lost to outflows as a result of the vigorous stellar activity. As the latter starts fading, the gas consumption levels off, as does the growth of new stellar mass, and the total mass within the region climbs up to roughly the level it had initially.

The behaviour is somewhat different within the smaller region ($R~<~2$~kpc). After an initial, nearly instantaneous star-formation burst, the SFR drops dramatically and maintains roughly the same level out to \mbox{$t~\approx~1$~Gyr}, at which point it increases significantly. There is a clear net gas mass inflow into the region ($\dot{M}~>~0$, orange curve), that may in fact lead to the formation of the inner bar.

In either region, the mass flow rate appears quasi-periodic (see sub-panels; grey curves), in agreement with \citet{li23b}, presumably as a result of the bar's rotation (cf. Fig.~\ref{fig:a2-phase}, right). A full Fourier analysis of the mass inflow to confirm or reject this suspicion is well beyond the scope of this paper and is therefore left for future work.

The delayed mass flow into the inner region suggests that the stars formed there should, on average, be younger compared to the average population of the outer region. In addition, given that the gas is flowing from the outer region and has likely to be enriched by earlier generations of stars, the mean metallicity of the stars within the inner region should, on average, be higher compared to the stars in the outer region.

\begin{figure*}
\centering
\includegraphics[width=0.33\textwidth]{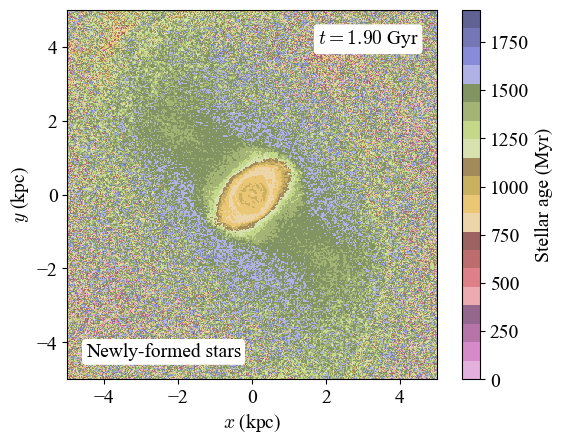}
\includegraphics[width=0.33\textwidth]{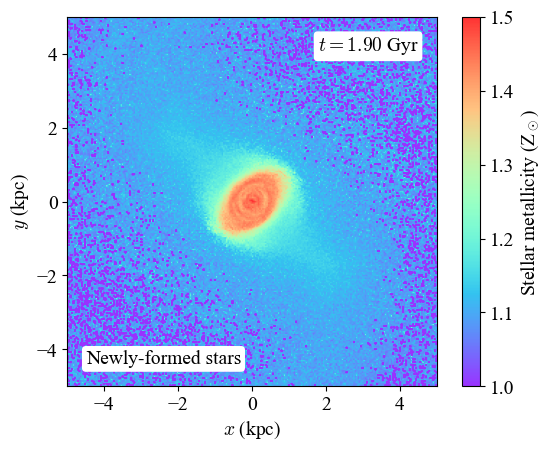}
\includegraphics[width=0.331\textwidth]{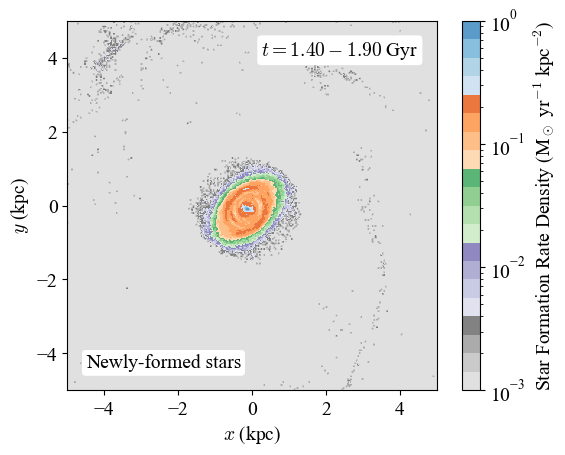}
\includegraphics[width=0.33\textwidth]{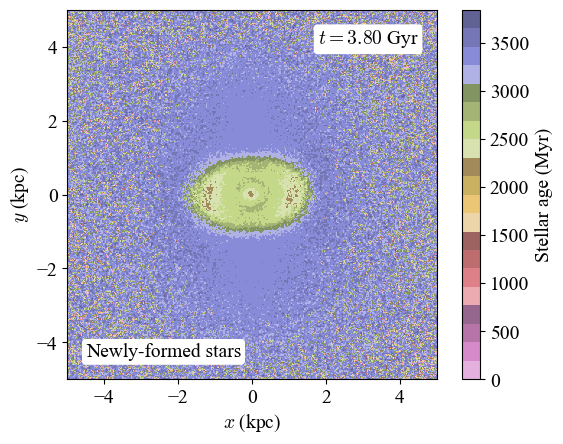}
\includegraphics[width=0.33\textwidth]{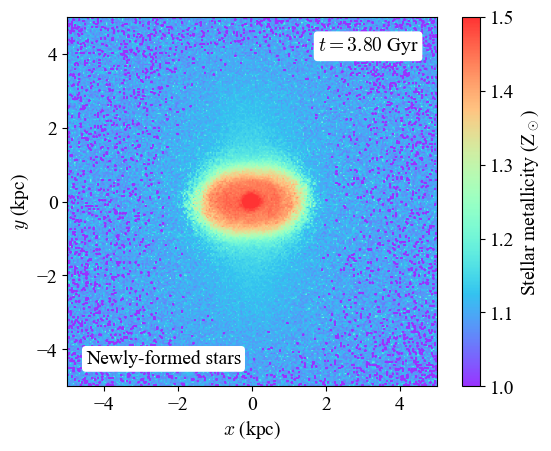}
\includegraphics[width=0.331\textwidth]{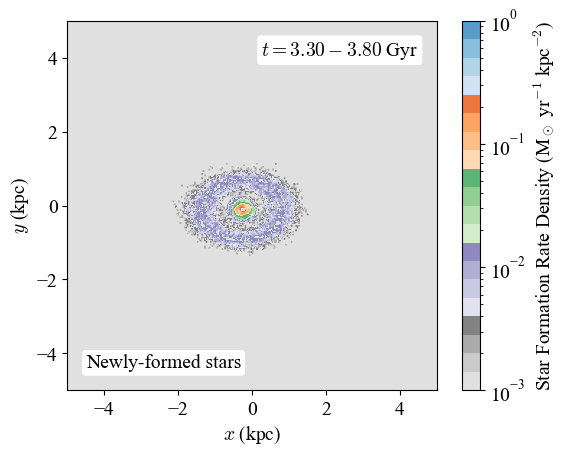}
\caption[  ]{ Mean stellar age (left), mean stellar metallicity (middle), and star-formation-rate density (SFRd; right) within the bar region ($R~\leq~5$~kpc) after \mbox{$t~\approx~1.9$~Gyr} (top) and 3.8~Gyr (bottom) of evolution. Note that the SFRd is calculated taking into account all the stars born in the last 500~Myr with respect to the given epoch. Clearly, the inner bar is younger compared to the outer bar at all times, but their ages both increase with time ({\em note the difference in the colour range between the middle and the bottom panels}). Note that the inner bar is more apparent here than it is in terms of its density distribution (Fig.~\ref{fig:nested_bars}), where it appears to have collapsed into a more circular mass concentration by the end of the simulation. Star formation is virtually only taking place within the inner bar region and decreasing rapidly with time, as can be seen by comparing the top and bottom panels in the last column. }
\label{fig:nested_bars_age_met_sfr}
\end{figure*}

In Fig.~\ref{fig:nested_bars_age_met_sfr} we display at two different epochs, \mbox{$t~=~1.9$~Gyr} and \mbox{$t~=~3.9$~Gyr}, the mean projected stellar age (left), the mean projected stellar metallicity (middle), and the star-formation rate density (in a time period of 500~Myr relative to the given epoch) within a region enclosed by \mbox{$(x,y)~\in~[-5,5]\times[-5,5]$~kpc}.
 
As suspected, we find that the stellar population of the inner bar is younger on average than that of the outer bar, implying that the inner bar forms after the outer bar, in agreement with \citet{woz15a}. Interestingly, this behaviour of bars formed due to an internal instability is the diametrically opposed behaviour to what is observed in the case where the bar formation is tidally induced \citep{sem24a}. Star formation is virtually only taking place within the inner bar region and decreasing rapidly with time, as can be seen by comparing the top and bottom panels in the last column, consistent with the results displayed in Fig.~\ref{fig:mass} (magenta curve).

\citet[][]{sem24a} note that in the case of NGC 1291, star formation can still happen outside the inner bar, after this was formed, and therefore ages of the inner bar can be still older than its surroundings \citep[see also][]{de-19a}. This is apparent in the top panel of Fig.~\ref{fig:nested_bars_age_met_sfr}, where an older (brown colour) nuclear stellar ring is surrounded by a younger structure (gold colour). The stellar age distribution is mirrored by the stellar metallicity distribution, as perhaps anticipated: younger stars are on average more enriched, and the more enriched stars occupy the regions closer to the centre. The presence of an older ring is also apparent in the metallicity map, visible as ring-like structure with a metallicity which is on average lower compared to its surroundings.

\begin{figure}
\centering
\includegraphics[width=1.02\columnwidth]{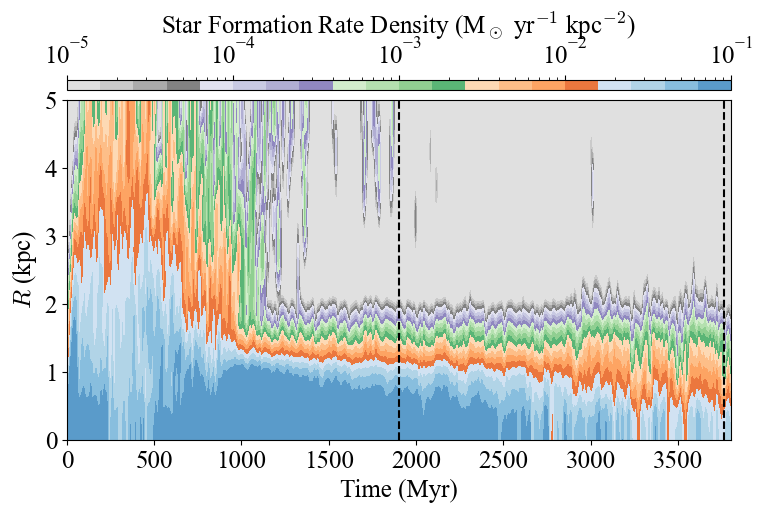}
\includegraphics[width=\columnwidth]{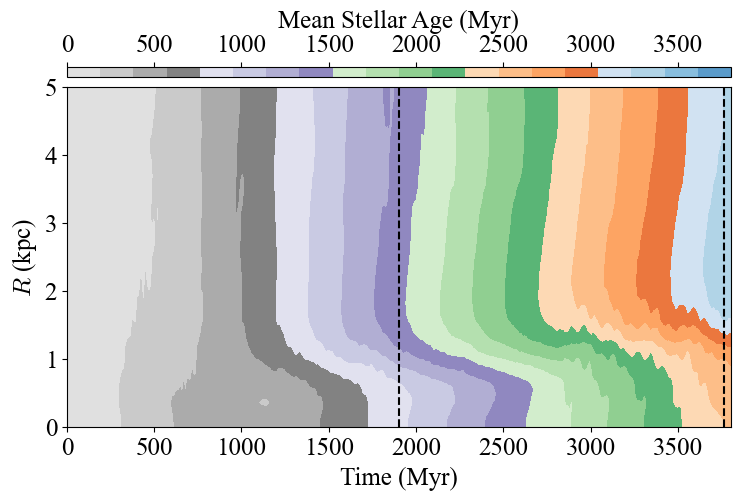}
\includegraphics[width=\columnwidth]{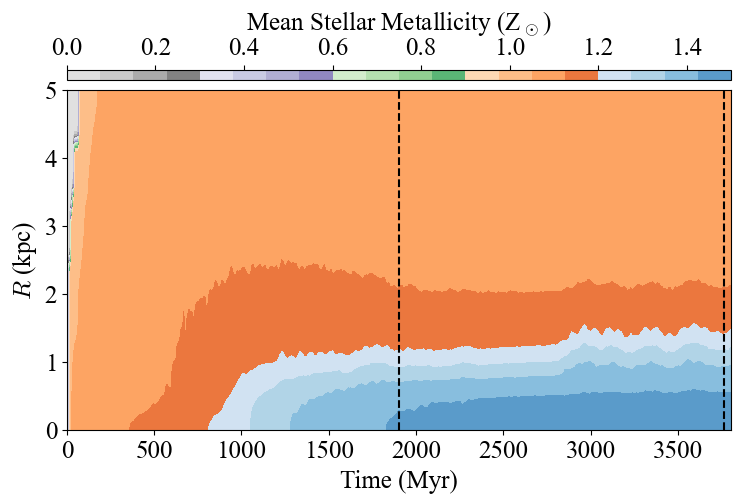}
\caption[  ]{ Top: Star-formation rate density within the bar region ($R~<~5$~kpc). Middle: Stellar age. Bottom: Stellar metallicity.  Although not always a good practice, in this case we consciously adopt the same colour scheme across all three panels to emphasise the connection between SFR on the one hand, and stellar age and stellar metallicity on the other. The vertical, dashed lines flag the epoch corresponding to the snapshots underlying Figs.~\ref{fig:nested_bars} and \ref{fig:nested_bars_age_met_sfr}. }
\label{fig:sfr_age_metal}
\end{figure}

A holistic view of the causal connection between SFR, stellar age and their enrichment is provided in Fig,~\ref{fig:sfr_age_metal}. It displays the evolution in space and time of the star-formation rate density (in units of \mbox{\Msun~yr$^{-1}$~kpc$^{-2}$}; top), the mean stellar age (in Myr; middle), and the mean stellar metallicity (in solar units; bottom).

It is apparent that the star-formation rate density decreases radially outwards at any given time, and also decreases with time at any given radius. Correspondingly, the mean stellar age at a given radius increases with time, and at any given time the mean stellar age decreases radially outwards. There is a clear trend separating younger from older populations at \mbox{$R~\approx~1$~kpc}, i.e.~between the inner and the outer bar regions, consistent with the age and metallicity distribution at the specific snapshots displayed in Fig.~\ref{fig:nested_bars_age_met_sfr}.

\subsection{Discussion}

Nested-bar systems (historically referred to as `double-barred' or S2B for short) have been known since \citet{de-75a,san79a}, and they have been studied in some detail both in observations \citep[][]{moi01a,erw04a} and simulations \citep[e.g.][]{deb07a,woz15a}. It has been suggested that even the MW may feature a nested bar \citep{ala01b,nam09a}. Roughly 20 percent of barred galaxies in the Local Universe feature a second, inner bar, and the frequency of S2B systems increases with stellar mass \citep{erw24a}. 

This type of system is not only interesting because of its exotic nature, but it is also of high relevance in the context of black-hole (BH) growth and active galactic nuclei (AGN) fuelling \citep[][]{shl89a,shl90a}. The basic idea is that the inner bar promotes the inflow of gas towards the centre beyond the radius that the outer bar usually does. It also has been suggested based on theoretical work that, if short-lived, dissolved inner bars may be the origin of bulges \cite[e.g.][]{du17a}.

The simulation presented here appears to support both of these beliefs. Indeed, the system undergoes a significant gas mass inflow towards the centre leading to an enhanced star formation, a younger and more enriched stellar population, compared to the surroundings. But the inner bar does not seem to be long-lived, i.e. over \mbox{$t~\gtrsim~1$~Gyr}, and rather appears to dissolve, yielding to the formation of a pseudo-spherical mass concentration (`bulge'), Whether this behaviour and the absence of a central black hole in the galaxy is related to a numerical aspect of our simulation such as the limited spatial resolution, or to a physical one such as the strength of the stellar feedback, is unclear at the moment.

Our simulation features both similarities and differences with respect to earlier, comparable simulations. For instance, the synthetic galaxy is bar unstable because it is baryon-dominated in the inner region \citep{fuj18a,bla23a}. This is in stark contrast to \citet{sah13b}, who find that a {\em dominant} halo, a hot disc, and no gas are needed for the system to spontaneously form a nested-bar structure. In comparison to \citet{sem24a}, in our simulation, the inner stellar bar appears perpendicular to the outer bar at nearly all times. The latter may be explained by the fact that the nested-bar nature of the galaxy in our simulation and \citet{sem24a}'s forms through different channels: In their case, it is tidally (i.e.~externally) triggered, while in our simulation it is internally triggered. In their case, the outer bar forms {\em after} the inner bar. Interestingly, in both simulations the galaxy shows the characteristic kinematic features of nested bars: the double quadrupole for mean $v_R$ and humps at the minor axis of the inner bar for $\sigma_z$ \citep[see also][]{de-08a}.

A notable difference with respect to earlier examples of simulated barred galaxies is that our synthetic galaxy displays what could be considered to be {\em three independent} bar components, i.e.~a triple-barred (S3B) galaxy, a beautiful example of which is the system NGC 6946 \citep{rom15a}. These authors provide evidence that a possible formation channel for such systems is a significant radial inflow of gas into the central region of a disc which already hosts a bar; this, together with gravitational instability and disc heating, then leads to the formation of the new, inner structures. Their interpretation is consistent with our results.

\section{Concluding Remarks: The case for controlled simulations} \label{sec:conc}

Controlled simulations of idealised galaxies sit between theoretical models and cosmological simulations, and are complementary to both. In other words, idealised galaxies are a natural bridge between the two, and accordingly we refer to our framework to perform this type of simulations as \nexus.

These simulations are very useful for isolating processes that are not easily discernible in a cosmological setting.
Their benefit is especially apparent when dealing with highly non-linear processes. Cosmological simulations have the advantage of treating the evolving hierarchy realistically, since mass assembly is largely driven by the dark matter. But once baryons are included, it becomes increasingly difficult to understand the role of interacting processes given the very large number of free parameters, made more difficult by inadequate numerical resolution.  In general, the best results have come from re-running a simulation of a specific ``local volume'' at  higher resolution \citep[e.g.][]{sor16a,ma17a} but even then, there are many competing processes to unwind. It is here where controlled simulations shine.

Famously, some of the earliest (restricted) N-body calculations \citep{too72a} revealed the emergence of bridges, tails and spiral-like features in galaxy-galaxy interactions.  Related phenomena include stellar shells  interleaved in radius that are due to an infalling satellite \citep{qui84a,bar92s}, now detected in cosmological simulations \citep{pop18a}. In a seminal paper, through controlled simulations, \citet{sel84a} showed how spiral instabilities are triggered by accretion and star formation, and are likely to be transient features of disc galaxies.

There are numerous examples of how controlled N-body simulations have led to new insights in galactic dynamics \citep[e.g.][]{ath92b,ath03l}. Typically, these manifestations are preceded by a strong theoretical basis, but not always \citep{bin08a}.
The discovery of two-dimensional \citep[2D; ][]{hoh71a} and three-dimensional \citep[3D; ][]{com90a} bar instabilities came from early N-body simulations of isolated discs. Later, it was shown that bars must slow down due to the exchange of energy through dynamical friction with a responsive dark matter halo \citep{deb98a}. A related process also makes a difference to the accretion of satellites in low-eccentricity orbits, and dynamical friction of these systems against the baryon disc can be much more important than against the dark halo \citep{wal96a}.

After \citet{sel02a} argued on theoretical grounds that stellar migration was possible, \citet{ros08v} realised the same behaviour in an isolated N-body disc, albeit with much higher rates of migration \citep[see also][]{min11a}. Inner and outer disc rings and resonances \citep{sel93a}, outer disc warps and flares \citep{bin92a,res98a} were first realised in controlled N-body simulations, as were other manifestations like disc heating due to disc-crossing satellites \citep{qui93a}. Some of these early experiments used rigid systems that did not conserve momentum and tended to exaggerate the energy exchange. Fully N-body simulations with live dark-matter halos established that the disc heating efficiency needs to be reduced by an order of magnitude \citep{hop08b}.

Controlled simulations continue to reveal new phenomena up to the present time.
When \citet{ant18b} discovered the remarkable `phase spiral' in the local disc, this was soon realised in a dozen controlled simulations \citep[e.g.][]{lap19a,kho19a,bla21e,hun21t} before being detected in cosmological simulations \citep[q.v.][]{gar22c}. \citet{aum16b} showed that the observed 3D stellar kinematic dispersions in the Milky Way result largely from the collective effects of scattering in both molecular clouds and spiral arms over cosmic time \citep[c.f.][]{ida93a}. Impressively, these processes are beginning to emerge in the best zoom-in simulations \citep{mcc24a}.
When a strong impulse (e.g.~disc-crossing massive satellite) warps the outer disc, this triggers a bending wave (corrugation) across the inner disc that {\it wraps up} with the disc's differential rotation \citep{bla21e}. Beyond the Milky Way, corrugations have since been seen in several disc galaxies with outer warps \citep{urr22a}.\\

We recognise that upcoming cosmological simulations will find ways to manufacture realistic gas-rich discs at very early cosmic times ($z\gtrsim 6$). This is likely to be a very complex and messy process. In part, our motivation for the new work is to provide a framework for simplifying what is seen in cosmological simulations with a view to understanding what happens with each added parameter (a {\em differential} approach). Nonetheless, synthetic galaxies at any epoch will continue to be highly idealised, as no method is yet able to fully accommodate e.g. magneto-hydrodynamical dynamos \citep[e.g.][]{bra05a,fed16b} and diffusion, supra-thermal particle and cosmic ray heating, dust and molecular processes, and so forth.\\

Our new framework is the ideal setting to understand the long-term relationship between the evolving ISM (governed mainly by the initial gas fraction) and stellar dynamical processes, referred to as {\em galactic ecology}. To this end, we will expand \nexus\ with additional (sub-grid) physics, which is readily available in \ramses, such as magneto-hydrodynamics  \citep[MHD;][]{Fromang2006} coupled to (highly) energetic particles \citep[`cosmic rays'; ][]{Dubois2016}. Cosmic rays are relevant since some theoretical work indicates that in some regimes they are able to significantly suppress star formation \citep[e.g.][]{Jubelgas2008,Pfrommer2017,Semenov2021} and drive colder outflows than what is predicted from pure supernova launching \citep[e.g.][]{Booth2013,Salem2014,Pakmor2016,Girichidis2018,Hopkins2020}.

In addition, next to the the expansion of our framework's capabilities in terms of the available halo models, as well as a more refined treatment of the gas chemistry, in upcoming work, we aim to couple our framework to the earliest disc systems emerging in the {\small VINTERGATAN} simulations that use, by design, the same star formation and feedback prescriptions \citep{age21l}. The ultimate goal is to create idealised analogues of galaxies that develop within a full cosmological context that can be run under controlled conditions.


\section*{Acknowledgements}
TTG acknowledges partial financial support from the Australian Research Council (ARC) through an Australian Laureate Fellowship awarded to JBH.
EV acknowledges support from an STFC Ernest Rutherford fellowship (ST/X004066/1).
OA acknowledges support from the Knut and Alice Wallenberg Foundation, the Swedish Research Council (grant 2019-04659), and the Swedish National Space Agency (SNSA Dnr 2023-00164).
CF acknowledges funding provided by the Australian Research Council (Discovery Project DP230102280), and the Australia-Germany Joint Research Cooperation Scheme (UA-DAAD). 
We further acknowledge high-performance computing resources provided by the Australian National Computational Infrastructure (grants ca64, ek9) and the Pawsey Supercomputing Centre (project~pawsey0810) in the framework of the National Computational Merit Allocation Scheme and the ANU Merit Allocation Scheme, by the Leibniz Rechenzentrum and by the Gauss Centre for Supercomputing (grants~pr32lo, pr48pi and GCS Large-scale project~10391).

Last, we are indebted to an insightful referee for a fair assessment of our work, which helped improve its overall presentation.\\

All figures and movie frames created with Matplotlib \citep{hun07a}. All animations assembled with {\small FFmpeg}.\footnote{\url{http://www.ffmpeg.org} }
This research has made use of NASA's Astrophysics Data System (ADS) Bibliographic Services.\footnote{\url{http://adsabs.harvard.edu} }

\section*{DATA AVAILABILITY}
The software and data underlying this article will be shared on reasonable request to the corresponding author.

\bibliographystyle{mnras} 
 \input{nexus.bbl} 

\bsp	
\label{lastpage}
\end{document}